\documentclass[12pt]{article}

\usepackage[margin=2cm]{geometry}
\usepackage[latin1]{inputenc}
\usepackage{graphicx}
\usepackage{fancyhdr}
\usepackage{amsfonts}
\usepackage{apacite}
\usepackage{color}

\author{
Ueli Rutishauser\\
Department Neural Systems and Coding,\\
Max Planck Institute for Brain Research\\ 
\texttt{urut@brain.mpg.de} \\
\and
Rodney J. Douglas\\
Institute of Neuroinformatics, University and ETH Zurich\\
\texttt{rjd@ini.phys.ethz.ch}\\
\and
Jean-Jacques Slotine \\
Nonlinear Systems Laboratory, Massachusetts Institute of Technology\\
\texttt{jjs@mit.edu} \\
}

\title{Collective stability of networks of winner-take-all circuits\footnote{This article has been published by MIT Press in Neural computation (2011). This is a pre-print version (as accepted). Please cite as Neural computation 23(3):735-773, 2011. The official version with final corrections is available at www.mitpressjournals.org/doi/abs/10.1162/NECO\_a\_00091}}

\newcommand{\bfx}{\mathbf{x}}

\newcommand{\bff}{\mathbf{f}}

\newcommand{\bfF}{\mathbf{F}}

\newcommand{\bfA}{\mathbf{A}}

\newcommand{\bfTheta}{\mathbf{\Theta}}

\begin{document}
\maketitle

\begin{abstract}
\begin{normalsize}
The neocortex has a remarkably uniform neuronal organization, suggesting that common principles of processing are employed throughout its extent. In particular, the patterns of connectivity observed in the superficial layers of the visual cortex are consistent with the recurrent excitation and inhibitory feedback required for cooperative-competitive circuits such as the soft winner-take-all (WTA). WTA circuits offer interesting computational properties such as selective amplification, signal restoration, and decision making. But, these properties depend on the signal gain derived from positive feedback, and so there is a critical trade-off between providing feedback strong enough to support the sophisticated computations, while maintaining overall circuit stability. The issue of stability is all the more intriguing when one considers that the WTAs are expected to be densely distributed through the superficial layers, and that they are at least partially interconnected. We consider the question of how to reason about stability in very large distributed networks of such circuits. We approach this problem by approximating the regular cortical architecture as many interconnected cooperative-competitive modules. We demonstrate that by properly understanding the behavior of this small computational module, one can reason over the stability and convergence of very large networks composed of these modules.  We obtain parameter ranges in which the WTA circuit operates in a high-gain regime, is stable, and can be aggregated arbitrarily to form large stable networks. We use nonlinear Contraction Theory to establish  conditions for stability in the fully nonlinear case, and verify these solutions using numerical simulations. The derived bounds allow modes of operation in which the WTA network is multi-stable and exhibits state-dependent persistent activities. Our approach is sufficiently general to reason systematically about the stability of any network, biological or technological, composed of networks of small modules that express competition through shared inhibition.
\end{normalsize}
\end{abstract}

\section{Introduction}

\begin{figure}
\centering
\includegraphics[angle=0,width=11cm]{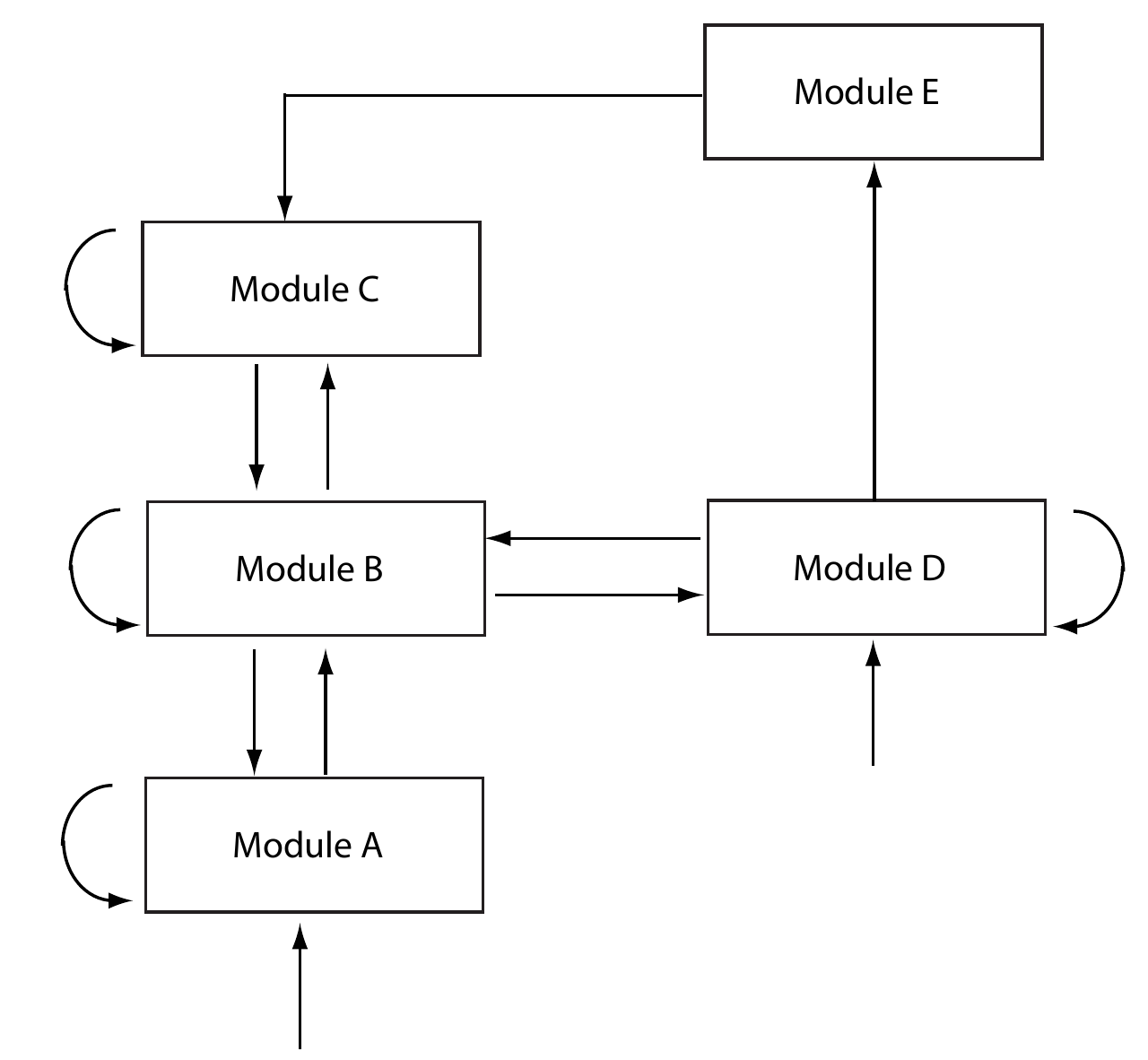}
\caption{Illustration of the problem. The two arrows at the bottom represent external input whereas all other connections are internal and excitatory. Shown is a recurrently connected system composed of 5 modules (each of which is a recurrent network). Given properties of the modules alone, can we guarantee the stability of the large connected system ? What constraints does each module have to observe for this to be true? }
\label{figProbIllustr}
\end{figure}

Large biological and artificial systems often consist of a highly interconnected assembly of components (Fig \ref{figProbIllustr}). The connectivity between these elements is often densely recurrent, resulting in various loops that differ in strength and time-constant \cite{Girard2008,Slotine01,Hopfield82,Amari77,Douglas95,Liu06}. This organization is true of the neocortex, where the statistics of connectivity between neurons indicate that recurrent connections are a fundamental feature of the cortical networks \cite{Douglas2004_neuronal,Binzegger04,Douglas95}. These recurrent connections are able to provide the excitatory and inhibitory feedback necessary for computations such as selective amplification, signal restoration, and decision making. But, this recurrence poses a challenge for the stability of a network \cite{Slotine01,Tegner02,Cohen83}. Connections may neither be too strong (leading to instability) or too weak (resulting in inactivity) for the network to function properly \cite{Koch99}.  In addition connections are continually changing as a function of learning, or are accumulated semi-randomly throughout development or evolution. How then, do these networks ensure stability?  Artificial neural networks can rely on their bounded (e.g. sigmoid) activation functions, but biological neurons do not usually enter saturation. Instead, their stability depends crucially on the balance between inhibition and excitation \cite{Hahnloser00,McCormick01}. In this paper we explore how the stability of such systems is achieved, not only because we wish to understand the biological case, but also because of our interest in building large neuromorphic electronic systems that emulate their biological counterparts \cite{Indiveri09}.

Reasoning about the computational ability as well as the stability of neural systems usually proceeds in a top-down fashion by considering the entire system as single entity able to enter many states (as for example in Hopfield networks \cite{Izhikevich07,Hopfield82,Hertz91}). Unfortunately, the number of states that must be considered grows exponentially with the size of the network, and so this approach quickly becomes intractable. For this reason stability analysis of large-scale simulations of the brain are proving difficult \cite{Izhikevich08,Ananthanarayanan09,Markram06}. 

We present an alternative approach, which uses bottom-up reasoning about the modules that constitute the network. The idea is that the stability of the modules should be conferred on the networks that they compose. Of course, simply combining several modules, each of which is stable in isolation, to form a larger system does not necessarily imply that the new system is stable \cite{Slotine01,Slotine03}. However, we explore the possibility that when the modules employ a certain kind of stability mechanism, then they are indeed able to confer stability also on the super-system in which they are embedded. We show that modules that achieve their own stability by observing constraints on their inhibitory/excitatory balance, can be stable alone as well as in combination. 

We have chosen to examine this problem in networks of WTA circuits \cite{YuilleGeiger03}, because these circuits are consistent with the observed neuroanatomical connections of cortex \cite{Douglas2004_neuronal,Binzegger04}. Moreover, the WTA is interesting because it can implement useful computational operations such as signal restoration, amplification, max-like winner selection (i.e. decision making) or filtering \cite{Maass00,Hahnloser99,Douglas2007_recurrent,YuilleGeiger03}. And, combining multiple WTAs in a systematic manner extends these possibilities further by allowing persistent activity and state-dependent operations \cite{RutishauserDouglas2009,Neftci08,Neftci_etal10}.

Typically, WTA networks operate in a high-gain regime in which their operation is non-linear (e.g. selective amplification). While the stability of a WTA can be analyzed by linearizing around the possible steady-states, rigorous analysis that takes the non-linearities into account is difficult using linear analysis tools \cite{strogatz94,Izhikevich07,Hahnloser98,Hahnloser03}.  Instead, we use nonlinear Contraction Analysis \cite{LohSlo98,Slotine03,Lohmiller2000} to investigate the stability of WTA networks. The concept of contraction is a generalization of stability analysis for linear systems, allowing Contraction Analysis \cite{LohSlo98} to be used for the analysis of circuits in the fully-non linear case, without making linearized approximations. 

A nonlinear time-varying system is said to be contracting if initial conditions or temporary disturbances are forgotten exponentially fast. Thus, any two initial conditions will result in the same system trajectory after exponentially fast transients. Importantly, the properties of contracting systems are preserved when they are combined to form a larger systems \cite{Slotine03}. Also, contraction allows parameter regimes which are not unduly restrictive. For instance, it can describe strong feedback loops; and, ranges of parameters can be found where the system is both contracting and operating in a highly non-linear regime. In addition, contraction analysis can deal with systems that are multi-stable (expressing several stable attractors or behaviors), where it guarantees exponentially fast convergence to one of the possible attractors. Such systems are capable of rich state-dependent computations, while at the same being contracting.
We have used Contraction Analysis to reason about the permissible kinds and strengths of connectivity within and between WTA modules embedded in a network. If the individual modules are contracting, then observing our constraints is sufficient to guarantee stability (boundedness) of a system composed of such modules. Thus, Contraction Analysis permits the derivation of simple bounds on the network parameters that will guarantee exponential convergence to equilibria in the fully non-linear case. This approach enables the systematic synthesis of large circuits, which are guaranteed to be stable if the set of bounds is observed. While we will demonstrate the feasibility of our approach in the case of WTA-type networks, our approach is not restricted to such networks. It can be applied as well to any simple non-linear circuit that is capable of non-linear computational operations.

\section{Results}
Our results are organized as follows. First, we introduce the basic organization of the WTA circuit. Second, we apply contraction theory to analyze the stability of networks of WTA circuits. We derive analytically the bounds on the parameters of the network that permit it to operate properly in either a soft-or hard WTA configuration. We conclude by performing numerical simulations to confirm that the analytical bounds are valid and not unnecessarily restrictive. 

\subsection{The winner-take all network}
\begin{figure}
\centering
\includegraphics[angle=0,width=11cm]{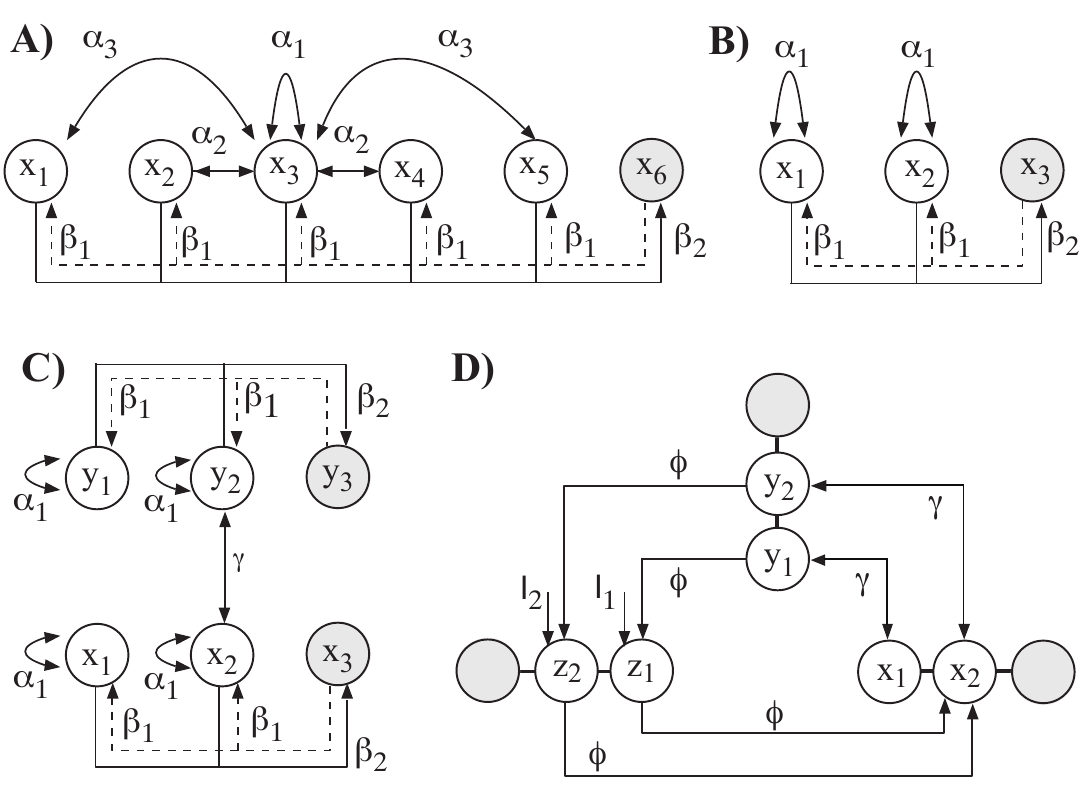}
\caption{Illustration of connectivity and notation. Excitatory and inhibitory connections are denoted by straight and dashed lines, respectively.  (A) Single WTA with all connections shown with respect to $x_3$. (B) Simplified version with $N=3$ units per map and $\alpha_2=0$ and $\alpha_3=0$. (C) Combination of two WTAs by symmetric bidirectional connection $\gamma$. (D) Network comprising 3 WTAs $x$, $y$ and $z$; connected by $\gamma$ and $\phi$ connections. The network has two states (each represented by one unit on maps $x$ and $y$) and two transition units $z_1$ and $z_2$. External input $I_1(t)$ or $I_2(t)$ to either $z_1$ or $z_2$ signals the arrival of a symbol to be processed by the network by executing a state dependent transition. If the network is in state 1, activation of $z_1$ initiates a transition from state 1 to 2. If the network is in state 2 and $z_2$ becomes active, the network remains in state 2 (a loop, see text). The local wiring on each WTA is not shown, but is equivalent to the connectivity of (B). 
}
\label{figConn}
\end{figure}

Each winner-take all (WTA) $\mathbf{x}$ consists of $N-1$ excitatory units $x_{1..N-1}$ and one inhibitory unit $x_N$ 
(See Fig \ref{figConn}A). Each excitatory unit receives recurrent input from itself ($\alpha_1$) and its neighbors ($\alpha_{2,3,...}$). For simplicity, only self-recurrence is considered here ($\alpha_{2,3,...}=0$), but similar arguments obtain when recurrence from neighboring units is included (see section \ref{sect:alpha2}). The inhibitory unit receives input from each excitatory unit with weight $\beta_2$, and projects to each excitatory unit with weight $\beta_1$. The dynamics of each unit are described by Eqs \ref{eq:recmapE} and \ref{eq:recmapI}. The firing rate activation function $f(x)$ is a non-saturating rectification non-linearity $max(0,x)$. The dynamics of this network, and in particular the  boundedness of its trajectories, depends on the balance of excitation and inhibition. 
 
\begin{equation}
\tau \dot{x}_i + G x_i = f( I_i + \alpha x_i - \beta_1 x_N - T_i)
\label{eq:recmapE}
\end{equation}

\begin{equation}
\tau \dot{x}_N + G x_N  = f( \beta_2\sum_{j=1}^{N-1} x_j - T_N)
\label{eq:recmapI}
\end{equation}

Where $I_i(t)$ is external input to unit $i$. All thresholds $T_i>0$ are constant and equal. $G>0$ is a constant that represents the load (conductance) and is assumed $G=1$, unless stated otherwise.  All parameters are positive: $\alpha>0, \beta_1>0, \beta_2>0$. We will refer to such a system either as a WTA or a "recurrent map" throughout the paper. ``Map`` will denote a WTA throughout, and not a discrete dynamical system.

\subsection{Combining several WTAs}
A single WTA network can implement some useful computational operations (see Introduction). However, more sophisticated computational operations can be achieved by combining several WTAs \cite{RutishauserDouglas2009} by sparse and selective connections between some of the excitatory units of the various WTAs. We consider two ways of combining WTAs: bidirectional and unidirectional. A bidirectional (and symmetric) connection establishes a recurrent connection between two WTAs. A unidirectional connection provides the activity of one WTA as input to a second WTA (feed-forward). The inhibitory units neither receive input from, nor do they project to, any other map. Thus, activity between maps is always excitatory (positive). This arrangement is motivated by the long range connections in cortex, which are predominantly excitatory \cite{Douglas2004_neuronal,Douglas95} (but they can contact both excitatory and inhibitory targets).
While long-range inhibitory projections in cortex exist as well, we focus exclusively on excitatory long-range connectivity in this paper.

These two kinds of connectivity are motivated by our previous finding that three WTAs connected by a combination of bi-and unidirectional connections are sufficient to implement state-dependent processing in the form of an automaton \cite{RutishauserDouglas2009}. An automaton consists of two components: states, and transitions between states. By connecting two maps bidirectionally, the network is able to maintain one region of persistent activity in the absence of external input, and this winning region represents the current state of the automaton. (State dependence is a form of memory and we thus refer to these localized regions of persistent activity as memory states.) Transition circuits allow the network to select a new winner, conditioned on the current state as well as an external input. The implementation of these transitions requires a third WTA (to select the most appropriate transition) as well as unidirectional connections between the maps that drive the transition (see below). In this paper we explore what constraints the presence of these additional connections poses on the stability of this and larger (more than three WTAs) networks.

First, consider two identical WTAs $\mathbf{x}$ and $\mathbf{y}$ (see Fig \ref{figConn}C). Each WTA consists of $N=3$ units (2 excitatory, 1 inhibitory). The only connection between the two networks is $\gamma$, which symmetrically (bidirectional) connects $x_2$ and $y_2$. Thus, this network can only represent one state.

The update equations for $x_2$ and $y_2$ thus become:
\begin{equation}
\tau \dot{x}_2 + G x_2 = f( I_2 + \alpha x_2 + \gamma y_2 - \beta_1 x_N - T)
\label{eq:recmapEcoupledT1}
\end{equation}
\begin{equation}
\tau \dot{y}_2 + G y_2 = f(  \alpha y_2 + \gamma x_2 - \beta_1 y_N - T)
\label{eq:recmapEcoupledT2}
\end{equation}


\label{sec:undirect}

Second, we consider unidirectional connections between WTAs. These are feed-forward connections between two maps: For example, when units on map $\mathbf{x}$ provide input to units on map $\mathbf{z}$. However, such feed-forward connections can result in (indirect) recurrence: For example, when map $\mathbf{z}$ in turn provides input to $\mathbf{x}$. Thus, analysis of unidirectional connections requires that we consider three maps $\mathbf{x}$,$\mathbf{y}$ and $\mathbf{z}$ simultaneously. The two maps $\mathbf{x}$ and $\mathbf{y}$ are connected bidirectionally as shown above, whereas $\mathbf{z}$ contains units that receive external input as well as input from $\mathbf{y}$ and also provide output to $\mathbf{x}$ (Fig \ref{figConn}D). In this way, strong enough activation of units on $\mathbf{z}$ can bias the ongoing competition in the network and thereby induce a switch to a new winner (so changing state). 

A given unit on $\mathbf{z}$ can either receive input from a different unit than it projects to (so providing a transition from one state to an other); or it can receive from and project to the same state. In Fig \ref{figConn}D, $z_1$ is an example of a unit that initiates a transition from state 1 to 2, whereas $z_2$ receives input from and projects to state 2. Thus, $z_2$ establishes an additional loop of recurrent feedback and is the more restrictive case when considering stability. 

Following Fig \ref{figConn}D, the dynamics of $x_1$ and $x_2$ become
\begin{equation}
\tau \dot{x}_1 + x_1 = f( I_1 + \alpha x_1 + \gamma y_1 - \beta_1 x_N - T)
\label{eq:recmapEcoupledTNX1}
\end{equation}
\begin{equation}
\tau \dot{x}_2 + x_2 = f( I_2 + \alpha x_2 + \gamma y_2 + \phi z_1 + \phi z_2 - \beta_1 x_N - T)
\label{eq:recmapEcoupledTNX2}
\end{equation}
\noindent and similarly for $y_1$, $y_2$.

The dynamics for the two new units $z_1$ and $z_2$ are
\begin{equation}
\tau \dot{z}_1 + z_1 = f( I_{TN1} + \alpha z_1 + \phi y_1 - \beta_1 z_N - T - T_{TN})
\label{eq:recmapEcoupledTNZ1}
\end{equation}
\begin{equation}
\tau \dot{z}_2 + z_2 = f( I_{TN2} + \alpha z_2 + \phi y_2 - \beta_1 z_N - T - T_{TN})
\label{eq:recmapEcoupledTNZ2}
\end{equation}
\noindent The equations for the other units of the system are equivalent to the standard WTA.

The term $T_{TNj}$ is an additional constant threshold for activation of the transition unit, so that in the absence of an external input $I_{TNj}$, the transition unit will remain inactive $z_j=0$. The external input $I_{TNi}$ can be used to selectively initiate a transition. An appropriate choice of the threshold $T_{TNj}$ will ensure that the transition unit $z_j$ is active only when both the external input $I_{TNi}>0$ and the input from the projecting map $y_j\phi>0$ are present. The activation of $z_j$ is thus state dependent, because it depends both on an external input as well as the current winner of the map. 

Now we will explore what constraints the presence of $\gamma>0$ and $\phi>0$ impose on stability. We will use Contraction Analysis to show that, if the single WTAs are contracting, $\gamma$ and $\phi$ can be used (with an upper bound) to arbitrarily combine WTAs without compromising the stability of the aggregate system. Since we base our arguments on contraction analysis, we will first introduce its basic concepts.

\subsection{Contraction Analysis}
\label{sec:contraction}
Essentially, a nonlinear time-varying dynamic system will be called
{\it contracting} if arbitrary initial conditions or temporary disturbances are
forgotten exponentially fast, i.e., if trajectories of the perturbed
system return to their unperturbed behavior with an exponential
convergence rate.  It turns out that relatively simple algebraic
conditions can be given for this stability-like property to be
verified, and that this property is preserved through basic system
combinations and aggregations.

A nonlinear contracting system has the following properties~\cite{LohSlo98,Lohmiller2000,Slotine03,WangSlo}
\begin{itemize}
	\item global exponential convergence and stability are guaranteed
	\item convergence rates can be explicitly computed as eigenvalues of well-defined Hermitian matrices
	\item combinations and aggregations of contracting systems are also contracting
	\item robustness to variations in dynamics can be easily quantified
\end{itemize}

Before stating the main contraction theorem, recall first the following.
The symmetric part of a matrix $\mathbf{A}$ is $\mathbf{A}_H=\frac{1}{2}(\mathbf{A}+\mathbf{A}^{*T})$. A
complex square matrix $\mathbf{A}$ is {\it Hermitian} if $\bfA^T = \bfA^*$ ,
where $^T$ denotes matrix transposition and $^*$ complex
conjugation. The {\it Hermitian part} $\bfA_H$ of any complex square matrix $\bfA$
is the Hermitian matrix $\frac{1}{2}(\bfA + \bfA^{*T})$ . All eigenvalues of a
Hermitian matrix are {\it real} numbers.  A Hermitian matrix ${\bf A}$
is said to be {\it positive definite} if all its eigenvalues are
strictly positive. This condition implies in turn that for any non-zero
real or complex vector ${\bf x}$, ${\bf x}^{*T}{\bf A}{\bf x}
> 0$. A Hermitian matrix ${\bf A}$ is called {\it negative definite} if $ - {\bf A}$ is positive definite.

A Hermitian matrix $\mathbf{A}(\mathbf{x},t)$ dependent on state or time will be called {\it
uniformly} positive definite if there exists a strictly positive constant such that for all states $\mathbf{x}$
and all $t \ge 0$ the eigenvalues of  $\mathbf{A}(\mathbf{x},t)$ remain larger than that constant.
A similar definition holds for uniform negative definiteness.

Consider now a general dynamical system in $\mathbb{R}^n$,
\begin{equation}
\label{eq:main}
	\dot \bfx = \bff(\bfx,t)
\end{equation}
with $\bff$ a smooth non-linear function. The central result of Contraction Analysis, derived in~\cite{LohSlo98} in both real and complex forms, can be stated as:

{\bf Theorem}
	Denote by $\frac{\partial \bff} {\partial \bfx}$ the Jacobian
        matrix of $\bff$ with respect to $\bfx$.  Assume that there exists a
        complex square matrix $\bfTheta(x,t)$ such that the Hermitian
        matrix $\bfTheta(x,t)^{*T}\bfTheta(x,t)$ is uniformly positive
        definite, and the Hermitian part ${\bf F}_H$ of the matrix 
	\[
        \bfF = \left(\dot\bfTheta + \bfTheta \frac{\partial \bff} {\partial \bfx}
        \right) \bfTheta^{-1} 
	\] 
        is uniformly negative definite. Then, all system trajectories converge
        exponentially to a single trajectory, with convergence rate
        $|\sup_{\bfx,t}\lambda_\mathrm{max}({\bf F}_H)|>0$. The system is said to
        be \emph{contracting}, $\bfF$ is called its \emph{generalized
        Jacobian}, and $\bfTheta(x,t)^{*T}\bfTheta(x,t)$ its contraction
        \emph{metric}. The contraction rate is the absolute value of the largest eigenvalue (closest to zero, although still negative) $\lambda = | \lambda_{max}\mathbf{F}_H |$. 

In the linear time-invariant case, a system is globally contracting if
and only if it is strictly stable, and $\bfF$ can be chosen as a
normal Jordan form of the system, with $\bfTheta$ a real matrix defining
the coordinate transformation to that
form~\cite{LohSlo98}. Alternatively, if the system is
diagonalizable, $\bfF$ can be chosen as the diagonal form of the
system, with $\bfTheta$ a complex matrix diagonalizing the system.
In that case, $\bfF_H$ is a diagonal matrix composed of the real parts
of the eigenvalues of the original system matrix.

Note that the activation function $f(x)=max(x,0)$ (see Eqs
\ref{eq:recmapE}-\ref{eq:recmapI}) is not continuously differentiable,
but it is continuous in both space and time, so that contraction
results can still be directly applied \cite{Lohmiller2000}. 
Furthermore, the activation function is piecewise linear with a
derivative of either $0$ or $1$. This simple property is exploited in the
following by inserting dummy terms $l_j$, which can either
be $0$ or $1$ according to the derivative of $f(x)$:
$l_j=\frac{d}{dx}f( x_j(t) )$. For a single WTA, there are a total of
$N$ dummy terms.



\subsection{Stability of a single WTA}
We begin the contraction analysis by considering a single WTA. The
conditions obtained in this section guarantee that the dynamics of the
single map converge exponentially to a single equilibrium point for a
given set of inputs. Actually, the WTA has several equilibrium points
(corresponding to each possible winner), but contraction analysis
shows that for a given input a particular equilibrium will be reached
exponentially fast, while all others are unstable. Thus, as long as
the network does not start out exactly at one of the unstable
equilibria (which is impossible in practice), it is guaranteed to
converge to the unique equilibrium point (the winner) determined by the given set of inputs. Our strategy is two-fold: first we show that
the WTA is contracting only if one of the excitatory units is active
(the "winner" in a hard-WTA configuration). Second, we show that in
the presence of multiple active excitatory units, the dynamics diverge
exponentially from the non-winning states.

Following section \ref{sec:contraction}, a system with Jacobian $\mathbf{J}$ is contracting if
\begin{equation}
\tau {\bf \Theta}\ {\bf J} \ {\bf \Theta}^{-1} \ < \ {\bf 0}
\label{eq:TW2}
\end{equation}
\noindent The Jacobian $\mathbf{J}$ has dimension $N$ and describes the dynamics of a single WTA, and $\mathbf{\Theta}$ is a transformation matrix (see section \ref{sec:contraction} and below). Using dummy terms $l_j$ as shown in the previous section, the Jacobian of the WTA is

\begin{equation}
\tau \mathbf{J} = \left[ 
\begin{array}{ccc}
l_1 \alpha-G & 0        & -\ l_1 \beta_1           \\
0        & l_2 \alpha-G & -\ l_2\beta_1           \\
l_3 \beta_2  & l_3 \beta_2  & -G                 \\
\end{array} 
\right]
\label{eq:Jmain}
\end{equation}

This WTA has two possible winners ($x_1$ or $x_2$) that are represented by $l_1=1, l_2=0$ or $l_1=0, l_2=1$, respectively ($l_3=1$ for both). Assuming the second unit is the winner, the Jacobian becomes

\begin{equation}
\tau \mathbf{J}_{W2} = \left[ 
\begin{array}{ccc}
-G 	 & 0            & 0           \\
0        & \alpha-G & -\ \beta_1           \\
 \beta_2  &  \beta_2  & -G                 \\
\end{array} 
\right]
\label{eq:JW2}
\end{equation}
\noindent Our approach consists in first finding a constant metric transformation
 $\mathbf{\Theta}$ describing the contraction properties
of the simple Jacobian~(\ref{eq:JW2}) for appropriate parameter ranges,
a process equivalent to standard linear stability analysis, and
then using the {\it same} metric transformation to assess the contraction
properties of the general nonlinear system.  

Let us first find ranges for the parameters $\alpha, \beta_1, \beta_2$
such that $\mathbf{J}_{W2}$ is contracting. This is the case if $\tau
{\bf \Theta}\ {\bf J}_{W2} \ {\bf \Theta}^{-1} \ < \ {\bf 0}$, where
$\mathbf{\Theta}$ defines a coordinate transform into a suitable
metric. The left hand-side is the generalized Jacobian
$\mathbf{F}=\mathbf{\Theta}\mathbf{J}_{W2}\mathbf{\Theta}^{-1}$ (see
section \ref{sec:contraction}). Based on the eigendecomposition
$\mathbf{J}_{W2} = \mathbf{Q} \mathbf{\Lambda} \mathbf{Q}^{-1}$, where
the columns of $\mathbf{Q}$ correspond to the eigenvectors of
$\mathbf{J}_{W2}$, define $\mathbf{\Theta} = \mathbf{Q}^{-1}$. This
transformation represents a change of basis which diagonalizes
$\mathbf{F}$ \cite{Horn85}. This choice of a constant invertible
$\bfTheta$ also implies that $\bfTheta^{*T}\bfTheta$ is positive
definite (since $\mathbf{x}^{*T} \bfTheta^{*T} \bfTheta \mathbf{x} =
|| \bfTheta \bfx ||^2$, $\forall {\bf x}$).

Using this transformation and assuming $G=1$, the Hermitian part of
$\mathbf{F}$ (Eq \ref{eq:TW2}) is negative definite if \footnote{\begin{normalsize}These
  solutions are derived by considering the eigenvalues of the
  Hermitian part of (\ref{eq:TW2}), which is diagonal and real, and
  then solving the system of inequalities $\lambda_{max}<0$.\end{normalsize}}
\begin{equation}
0<\alpha<2 \sqrt{\beta_1 \beta_2 }
\label{eq:limAlpha}
\end{equation}
\begin{equation}0<\beta_2
\label{eq:limBeta2}
\end{equation}
\begin{equation}
0<\beta_1\beta_2<1
\label{eq:limBeta1}
\end{equation}
\noindent Note that these conditions systematically relate $\alpha$ to
the inhibitory loop gain $\beta_1 \beta_2$, and also permit $\alpha>1$
(see below for discussion).

The above conditions guarantee contraction for the cases where
inhibition ($l_3=1$) and one excitatory unit are active (here $l_2=1$
and $l_1=0$ but the same bounds are valid for $l_2=0$ and
$l_1=1$). The next key step is to use the same metric to study {\it arbitrary}
terms $l_{2,3}$ and $l_1=0$, so as to show that the system is contracting for
all combinations of $l_{2,3}$, except the combinations from which we want
the system to be exponentially diverging.  In the same metric
$\bfTheta$ and using the Jacobian Eq \ref{eq:Jmain} with $l_1=0$ the
Hermitian part of $\mathbf{F}$ becomes (with $i^2 = - 1$) 
\begin{equation}
\mathbf{F_H} = \left[ 
\begin{array}{ccc}
-1 &   0  & 0           \\
0  & -1+\frac{1}{2}\alpha l_2 & -\frac{(2 i \beta_1 \beta_2+\alpha (-i \alpha+\sqrt{-\alpha^2+4 \beta_1 \beta_2})) (l_2-l_3)}{2 \sqrt{-\alpha^2+4 \beta_1 \beta_2}}           \\
0  &  -\frac{(-2 i \beta_1 \beta_2+\alpha (i \alpha+\sqrt{-\alpha^2+4 \beta_1 \beta_2})) (l_2-l_3)}{2 \sqrt{-\alpha^2+4 \beta_1 \beta_2}}  & -1+\frac{1}{2}\alpha l_2        \\
\end{array} 
\right]
\label{eq:Fhermit1}
\end{equation}
\noindent Note that Eq (\ref{eq:Fhermit1}) was simplified assuming the bound given in Eq \ref{eq:limAlpha}. We require $\mathbf{F_H}<0$. A matrix of the form 
$\left[ \begin{array}{cc}
\lambda_1 & r  \\
r^{*}     & \lambda_2 \\
\end{array} 
\right]$ 
is negative definite if $\lambda_i<0$ and $|r|^2< \lambda_1 \lambda_2$ \cite{WangSlo}. For (\ref{eq:Fhermit1}), this results in
\begin{equation}
\frac{ (\beta_1 \beta_2 (l_2-l_3))^2}{-\alpha^2+4 \beta_1 \beta_2} < (-1 + \frac{1}{2}\alpha l_2)^2
\label{eq:Fhermit2}
\end{equation}
\noindent The bounds (\ref{eq:limAlpha})-(\ref{eq:limBeta1}) on the
parameters satisfy this condition whenever $l_2=l_3$ and $l_2=0,
l_3=1$.  As expected, for the case $l_2=1, l_3=0$ (only excitation
active) the system is not contracting for $\alpha>1$. Rather, we
require that in this case the system is exponentially diverging, as we
detail below.

Next, we consider the full Jacobian (Eq \ref{eq:Jmain}) with all
$l_{1,2,3}=1$. For the network to be a hard-WTA, we require that this
configuration is exponentially diverging. The dynamics of interest are
the excitatory units, so that, following \cite{PhamSlotine07}, the
system is exponentially diverging away from this state if
\begin{equation}
\mathbf{V} \mathbf{J} \mathbf{V}^T > \mathbf{0}
\label{eq:VJVpos}
\end{equation}
 \noindent where ${\bf V}$ is the projection matrix
\begin{equation}
\mathbf{V} = \left[ 
\begin{array}{ccc}
\alpha 	 & 0      & -\ \beta_1           \\
0        & \alpha & -\ \beta_1           \\
\end{array} 
\right]
\label{eq:V1}
\end{equation}
\noindent The constraint (\ref{eq:VJVpos}) assures that the system diverges from certain invariant subspaces where $\mathbf{V}\mathbf{x}$ is constant. For $\mathbf{V}$ as shown in (\ref{eq:V1}),
$\mathbf{V}\mathbf{x}=\left[ \begin{array}{c}
\alpha x_1  - \beta_1 x_3 \\
\alpha x_2  - \beta_1 x_3 \\
\end{array} 
\right]$. Each row represents one excitatory unit. 
If condition (\ref{eq:VJVpos}) is satisfied, the network is guaranteed to diverge exponentially away from this equilibrium.

Condition (\ref{eq:VJVpos}) is satisfied (for $G=1$) if 
\begin{equation}
1<\alpha
\label{eq:limVJVpos1}
\end{equation}
\begin{equation}
0<\beta_1
\label{eq:limVJVpos2}
\end{equation}
\begin{equation}
0<\beta_1 \beta_2 < (1 - \frac{1}{\alpha})(\beta_1^2 + \frac{\alpha^2}{2})
\label{eq:limVJVpos3}
\end{equation}

The above conditions were derived based on the system of inequalities $\lambda_{min}>0$ given by the eigenvalues of the Hermitian part of the left-hand side of (\ref{eq:VJVpos}). The same calculation using instead $l_{1,2}=1$ and $l_3=0$ (excitation, but no inhibition) results in the same bounds for exponential divergence from the state of no inhibition.

Combining (i) conditions (\ref{eq:limAlpha})-(\ref{eq:limBeta1}) for exponential convergence to the winner state and (ii) conditions (\ref{eq:limVJVpos1})-(\ref{eq:limVJVpos3}) for exponential divergence from the non-winning and the excitation-only states, yields
\begin{equation}
1<\alpha<2 \sqrt{\beta_1 \beta_2 }
\label{eq:limAlphaDiv}
\label{eq:keyResultBegin}
\end{equation}
\begin{equation}
\frac{1}{4}<\beta_1\beta_2<1
\end{equation}
\begin{equation}
\beta_1 \beta_2 < (1 - \frac{1}{\alpha})(\beta_1^2 + \frac{\alpha^2}{2})
\label{eq:keyResultEnd}
\end{equation}

\noindent Note the two key components: the excitatory gain $\alpha$ and the inhibitory gain $\beta_1\beta_2$. 
The above conditions establish lower and upper bounds on the parameters for global exponential convergence to a unique winner for a given set of inputs. 

\begin{figure}
\centering
\includegraphics[angle=0,width=11cm]{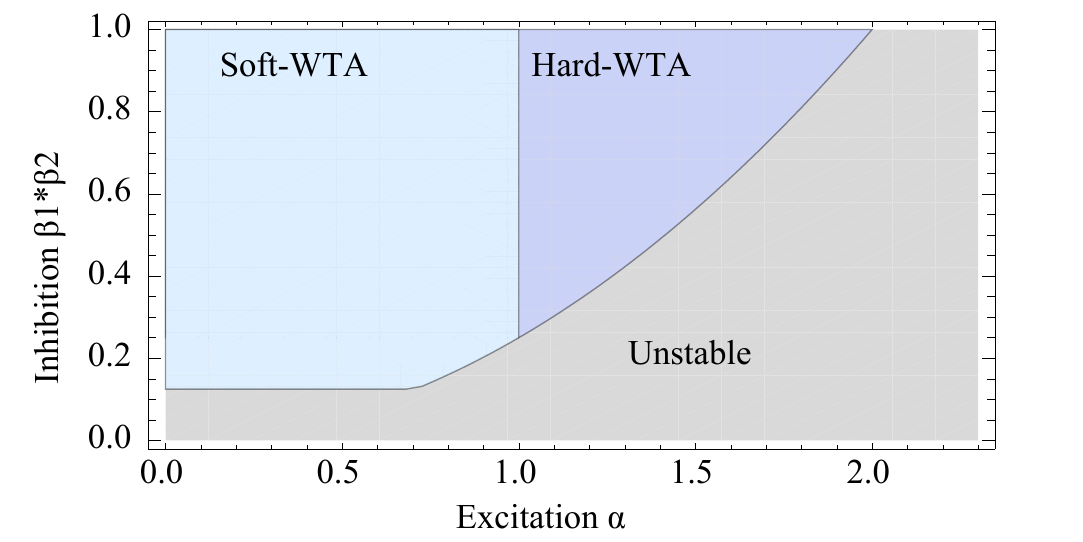}
\caption{Illustration of the different operating conditions as a function of the excitatory strength $\alpha$ and the inhibitory loop gain $\beta_1\beta_2$. Note the transition from a soft-WTA to hard-WTA at $\alpha=1$.}
\label{figGraphicalIneq}
\end{figure}

Under these constraints (in particular on the excitatory loop strength $\alpha$) the system is globally convergent yet always selects a winner. The system does not depend on saturation to acquire this stability. Also, the constraints guarantee that the system does not oscillate, apart from transient oscillations during convergence.
This has been established by demonstrating that the system is either contracting or exponentially diverging for any subset of the dummy terms $l_{1,2,3}$.
Note that the system is contracting in the {\it same} metric $\bfTheta$ for all contracting subsets. While we defined the metric $\bfTheta$ for a particular  winner, the same constraints result from defining a similar $\bfTheta$ for any of the other possible winners. 
Similar conditions can be derived for conditions where the winner is represented by multiple active units such as when a "bump of activity" is introduced by adding excitatory nearest-neighbor connections $\alpha_2$ \cite{RutishauserDouglas2009,Douglas2007_recurrent} (see section \ref{sect:alpha2}). Numerically, these ranges permit a wide range of parameters. For example, for $\beta_1=3$ and $\beta_2=0.3$, $1<\alpha<1.89$. Under these conditions, the system operates in a highly non-linear regime (where the loop gain can be up to ~50!).

The analysis above focused on the regime where $\alpha>G$ (with $G=1$). In this mode, the system acts as a highly non-linear WTA, always selecting a binary winner. What if the system operates in $\alpha<1$ ? In this configuration, the winner unit is still contracting (Eq \ref{eq:limAlpha}).

What happens when all units ($l_{1,2,3}=1$) are active and $\alpha<1$? Defining $\mathbf{\Theta}$ based on the Jacobian $\mathbf{J}$ with all units on and solving $\mathbf{\Theta}\mathbf{J}\mathbf{\Theta}^{-1}<\mathbf{0}$, we find that this system is contracting for $\alpha<1$. The system where all excitatory units are active is thus contracting under this condition, implying that the system is in a "soft-WTA" configuration. While the system still selects a winning unit, the activity of the loosing unit is not completely suppressed. Also note that in this configuration, no persistent activity in the absence of external input is possible. A graphical illustration of both modes of operation is shown in Fig \ref{figGraphicalIneq}.

Finally, note that the time-constant $\tau$ was assumed to be equal
for all units. Note that, in this case, the numerical value of $\tau$
does not influence the bounds (since $\tau>0$ multiplies the entire
Jacobian, see Eq \ref{eq:TW2}). Similar conditions can be derived for
conditions where the time-constants are not equal (see Appendix C), in
which case only the ratio of the time-constants is relevant.

\subsection{Stability of single WTA of arbitrary size}
\begin{figure}
\centering
\includegraphics[angle=0,width=11cm]{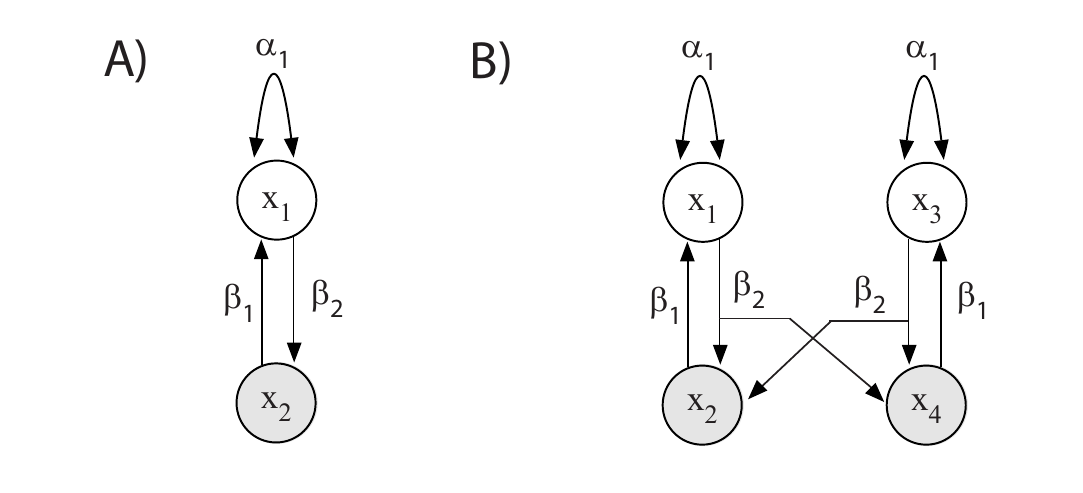}
\caption{Illustration of combining a simple WTA to form a bigger WTA . White units are excitatory, gray units inhibitory. (A) Simple map consisting of one excitatory unit $x_1$ and one inhibitory unit $x_2$. (B) Combining two identical copies of the map shown in (A) by providing excitatory input $\beta_2$ from each excitatory unit to both inhibitory units. This new map, consisting of two excitatory units, is functionally equivalent to a WTA with two excitatory and one inhibitory units. The two inhibitory units $x_2$ and $x_4$ have, at all times, the same level of activity. The same principal can be extended to form maps of arbitrary size.}
\label{figSingleMap}
\end{figure}

Can this analysis be extended to maps of arbitrary size? While the approach in the previous section can be applied to maps of any size, an alternative approach is to first define contraction for a map consisting only of a single excitatory and inhibitory unit and then extend it recursively by one unit at a time, while showing that this extension does not change the contraction properties. This approach is illustrated in Fig \ref{figSingleMap}.

The most simple map consists of one excitatory and one inhibitory unit (Fig \ref{figSingleMap}A) . While there is no competition between different inputs, this map otherwise preserves all the properties of a WTA (such as non-linear amplification of the input). The Jacobian of this map is:

\begin{equation}
\tau \mathbf{A} = \left[ 
\begin{array}{ccc}
\l_1 \alpha-G & -\ \l_1 \beta_1   \\
\l_2 \beta_2  & -G           \\
\end{array} 
\right]
\label{eq:Js1}
\end{equation}

This system (Fig \ref{figSingleMap}A) is contracting if the conditions shown in Eqs  \ref{eq:keyResultBegin}-\ref{eq:keyResultEnd} for the parameters $\alpha, \beta_1, \beta_2$ hold. The approach used to derive the bounds is equivalent to the one described above: first, define a $\mathbf{\Theta} = \mathbf{Q}^{-1}$, where $\mathbf{Q}$ is based on the eigendecomposition of $\mathbf{A}$ with $l_{1,2}=1$. Then, define the valid parameter ranges based on Eq \ref{eq:TW2}. The same permissible parameters result (see Eqs \ref{eq:limAlpha}, \ref{eq:limBeta1}, \ref{eq:limBeta2}).

Combining two such maps by feeding excitatory input to both inhibitory neurons by both excitatory neurons leads to a WTA with two excitatory units (Fig \ref{figSingleMap}B). This map is equivalent to the map shown previously, except for that it contains two inhibitory neurons. These are, however, functionally equivalent (their activity and derivatives are the same at all points of time). Thus the behavior of both systems will be equivalent. The Jacobian of the combined system is:

\begin{equation}
\tau \mathbf{J} = \left[ 
\begin{array}{ccc}
\mathbf{A}_{1} & \mathbf{G_1}   \\
\mathbf{G_2}  & \mathbf{A}_{2}           \\
\end{array} 
\right]
\label{eq:Jrecursive2}
\end{equation}

where 

\begin{equation}
\tau \mathbf{G} = \left[ 
\begin{array}{ccc}
0 & 0   \\
l_2 \beta_2  & 0 \\
\end{array} 
\right]
\end{equation}

and $\mathbf{A}_{1,2} = \mathbf{A}$ after adjusting the $l_i$ terms appropriately ($l_{1,2}$ and $l_{3,4}$ for $\mathbf{A}_{1}$ and $\mathbf{A}_{2}$, respectively). Similarly, $\mathbf{G}_{1,2}=\mathbf{G}$ for $l_2$ and $l_4$, respectively. 
Note that combining the two systems in this way adds only two (strictly positive) terms to the equations describing the dynamics of the inhibitory neurons. Thus, inhibition in this new system can only be larger compared to the smaller system. Thus, if the smaller system is contracting (as shown above), the combined system must also be contracting (shown in the next section).

Defining a metric $\mathbf{\Theta}$ based on the eigendecomposition of $\mathbf{J}$ for either $l_{1,2,4}=1$ and $l_3=0$ or $l_{1,3,4}=1$ and $l_2=0$ and then solving

\begin{equation}
\tau {\bf \Theta}\ {\bf J} \ {\bf \Theta}^{-1} \ < \ {\bf 0}
\label{eq:JrecursiveF}
\end{equation}
\noindent results in the same constraints for the system to be contracting (see Eqs \ref{eq:limAlpha}, \ref{eq:limBeta1}, \ref{eq:limBeta2}). 

This result can be generalized so that it is valid for adding one unit to a map that is already contracting. This can be seen directly by considering the eigenvalues of the Hermitian part of $\mathbf{F}=\mathbf{\Theta}\mathbf{J}\mathbf{\Theta}^{-1}$, defined either for a system with $n$ units or $n+1$ units. A system with $n=1$ units has Jacobian $\mathbf{A}_1$ and is contracting as shown previously. The condition for it to be stable $\mathbf{F_s}<\mathbf{0}$ requires (for the real part only)
\begin{equation}
\frac{1}{2}(-2+\alpha\pm\sqrt{\alpha^2-4\beta_1\beta_2})<0
\label{eq:eigValsN1}
\end{equation}
\noindent A system with $n=2$ units has Jacobian $\mathbf{J}$ (Eq \ref{eq:Jrecursive2}) and is stable if (\ref{eq:JrecursiveF}) holds. This requires
\begin{equation}
\frac{1}{2}(-2+\alpha\pm\sqrt{\alpha^2-4\beta_1\beta_2})<0
\label{eq:eigValsN2}
\end{equation}
\noindent Comparing Eqs \ref{eq:eigValsN1} and \ref{eq:eigValsN2} reveals that adding a unit $n+1$ to a system of $n$ units does not change the conditions for contraction to a single winner. Thus, if the recurrent map consisting of $n$ excitatory unit is contracting the system of $n+1$ units is also contracting. By recursion this proof can be applied to maps of arbitrary size.

What if multiple units on the map are active? Above conditions show that a single winner is contracting on an arbitrary sized map. In a hard-WTA configuration, the system should always emerge with a single winner. We have previously shown that our system has this property when $\alpha>1$ (see Eq \ref{eq:VJVpos}). Here, we extend this argument to maps of arbitrary size. Note that only the $l_i$ for the excitatory units can be switched inactive. All inhibitory neurons (since they all represent the same signal) are always $l_i=1$. 

Here, we start with a system that has $n=2$ units (since a system of $n=1$ does not have competition). The goal is to find conditions that enforce a single winner for $n=n+1$ units. For the $n=2$ system ($\mathbf{J}$ with all $l_{1,2,3,4}=1$), enforcing $\mathbf{V} \mathbf{J} \mathbf{V}^T > \mathbf{0}$ (see Eq \ref{eq:VJVpos}) with
\begin{equation}
\mathbf{V} = \left[ 
\begin{array}{cccc}
\alpha 	 & -\beta_1   & 0 &  -\beta_1           \\
0        & -\beta_1   & \alpha & -\ \beta_1           \\
\end{array} 
\right]
\label{eq:VtermN2}
\end{equation}
\noindent gives conditions for this configuration (both units on, i.e. $l_1=l_3=1$) to be exponentially unstable (thus converging to an other subset of the $l_i$ terms). Similar to (\ref{eq:V1}), the system diverges from invariant subspaces where $\mathbf{V}x$ is constant. For the projection (\ref{eq:VtermN2}), $\alpha x_1  - \beta_1 x_2 - \beta_1 x_4 =0$ and $\alpha x_3  - \beta_1 x_2 - \beta_1 x_4 =0$ defines the equilibrium. If condition (\ref{eq:VJVpos}) is satisfied, the network is guaranteed to diverge exponentially away from this equilibrium.

The eigenvalues of the Hermitian part of this system (same as for Eq \ref{eq:VJVpos}) are uniformly positive if the following two conditions hold
\begin{equation}
-\alpha^2+\alpha^3 > 0, -\alpha^2+\alpha^3-\beta_1^2+2 \alpha \beta_1^2-4 \alpha \beta_1 \beta_2 > 0
\label{eq:eigValsV2}
\end{equation}
\noindent Note that any solution requires $\alpha>1$ (solutions are shown in (\ref{eq:limVJVpos1})-(\ref{eq:limVJVpos3})). This condition thus shows that any two simultaneously active units can not be contracting if $\alpha>1$.

For the 3 unit system, applying (\ref{eq:Jrecursive2}) recursively results in the Jacobian
\begin{equation}
\tau \mathbf{J} = \left[ 
\begin{array}{ccc}
\mathbf{A}_{1} & \mathbf{G_1}   & \mathbf{G_1}  \\
\mathbf{G_2}   & \mathbf{A}_{2} & \mathbf{G_2}          \\
\mathbf{G_3}   & \mathbf{G_3}   & \mathbf{A}_{3}        \\
\end{array} 
\right]
\label{eq:Jrecursive3}
\end{equation}
\noindent Applying an appropriate $\mathbf{V}$ constructed in analog to (\ref{eq:VtermN2}) shows that $\mathbf{V} \mathbf{J} \mathbf{V}^T > \mathbf{0}$ for this system if
\begin{equation}
-\alpha^2+\alpha^3 > 0, -\alpha^2+\alpha^3-6\beta_1^2+3 \alpha \beta_1^2-9 \alpha \beta_1 \beta_2 > 0
\label{eq:eigValsV3}
\end{equation}
\noindent Note that a sufficient solution continues to require $\alpha>1$. We have thus shown, under $\alpha>1$ that a system with $n=2$ as well as $n=3$ can only have one active unit. By recursion, the same argument can be used to show that any system $n=n+1$ can not have a subset of $i$ units (where $1<i<=n$)  active. Any such system is thus always converging to a single winner. Any such system will have these properties if the parameters are within the ranges shown in Eqs \ref{eq:limVJVpos1}, \ref{eq:limVJVpos2}, \ref{eq:limVJVpos3}) hold.

For purposes of this proof, we used additional inhibitory units (one for each excitatory unit). Note that this arrangement is for mathematical convenience only: In an implemented system these units can be collapsed to one unit only (or several to implement local inhibition). Collapsing all units to one does not change the dynamics of the system, because all inhibitory units have the same activity (and its derivatives) at all times.

\subsubsection{Example}

This example will show how to apply the approach outlined above in order to calculate the permissible range of parameters for a toy recurrent map consisting of one excitatory and one inhibitory unit (Fig \ref{figSingleMap}A), whose Jacobian is  $\mathbf{A}$ (Eq \ref{eq:Js1}) with $l_{1,2}=1$. Our intention is to illustrate in detail the procedural aspects involved in the calculation.

First construct $\mathbf{Q}$ based on the eigenvectors of $\mathbf{A}$ and then set

\begin{equation}
\mathbf{\Theta} = \mathbf{Q}^{-1} = \left[ 
\begin{array}{cc}
 -\frac{{\beta_2}}{\sqrt{\alpha^2-4 {\beta_1} {\beta_2}}} & \frac{1}{2} \left(1+\frac{\alpha}{\sqrt{\alpha^2-4 {\beta_1} {\beta_2}}}\right) \\
 \frac{{\beta_2}}{\sqrt{\alpha^2-4 {\beta_1} {\beta_2}}} & \frac{1}{2}-\frac{\alpha}{2 \sqrt{\alpha^2-4 {\beta_1} {\beta_2}}}
\end{array}
\right]
\end{equation}
\noindent Then, transforming $\mathbf{A}$ using $\mathbf{\Theta}$ results in the generalized Jacobian
\begin{equation}
\mathbf{F} = \tau \mathbf{\Theta} \ \mathbf{A} \ \mathbf{\Theta}^{-1} = \left[  
\begin{array}{cc}
 \frac{1}{2} \left(-2+\alpha-\sqrt{\alpha^2-4 \beta_1 \beta_2}\right) & 0 \\
 0 & \frac{1}{2} \left(-2 + \alpha + \sqrt{\alpha^2-4 \beta_1 \beta_2}\right)
\end{array}
\right]
\label{eq:numericThetaApply}
\end{equation}

Due to the choice of the metric $\Theta$, only terms on the diagonal remain. The network is contracting if the Hermitian part $\mathbf{F}_H=\frac{1}{2}(\mathbf{F}+\mathbf{F}^{*T})$ (\ref{eq:numericThetaApply}) is negative definite. A sufficient condition for this to be the case is $Re( \lambda_{min}(\mathbf{F}_s))<0$. Solving this system of inequalities results in the conditions shown in (\ref{eq:limAlpha}, \ref{eq:limBeta2}, \ref{eq:limBeta1}).

\subsubsection{Comparison with numerical simulations}
\begin{figure}
\centering
\includegraphics[angle=0,width=11cm]{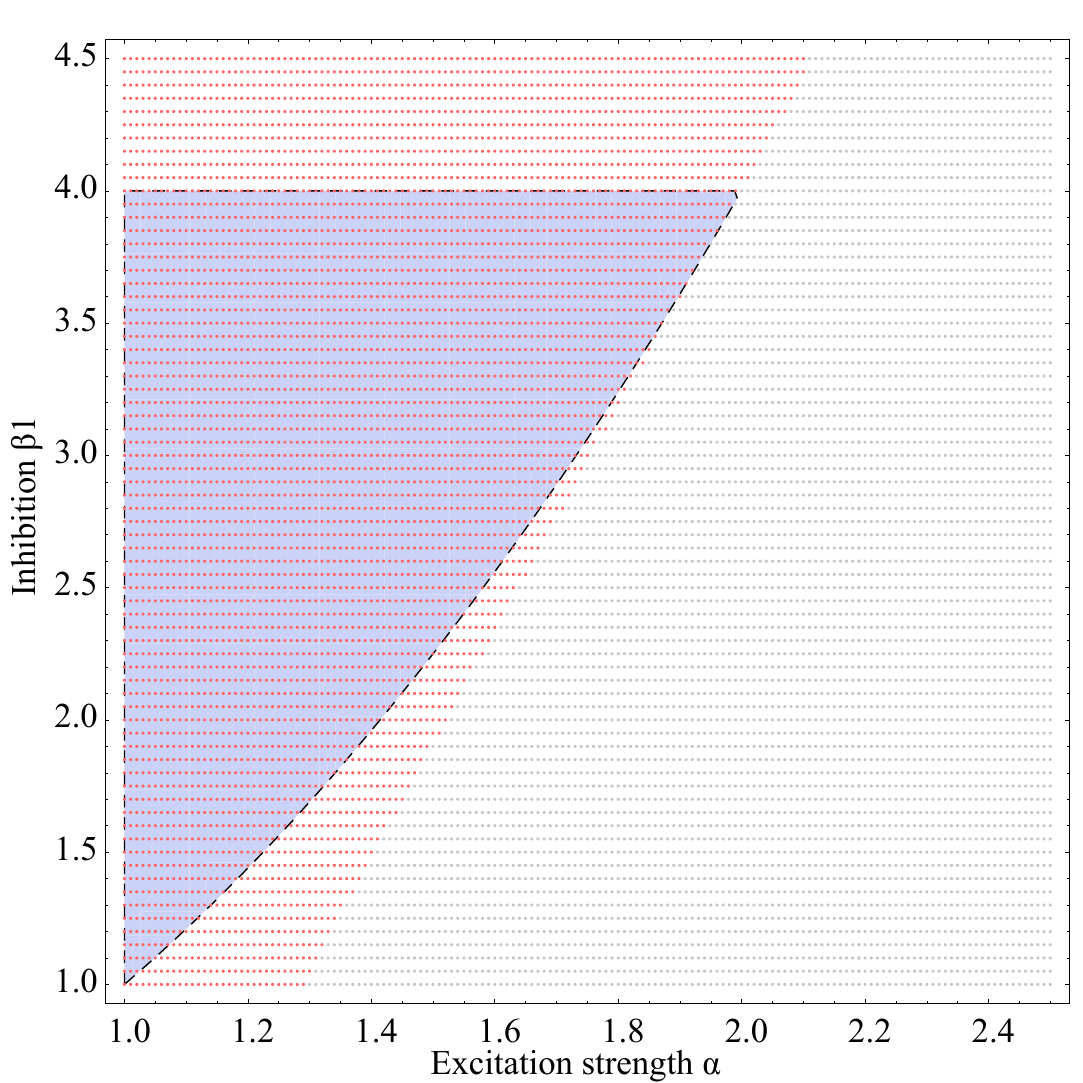}
\caption{Comparison of the permissible parameter range derived analytically, with that obtained by simulation. Results are shown as a function of $\alpha$ and $\beta_1$, while holding $\beta_2=0.25$ (constant). The region of contraction is indicated in light blue. The upper boundary is $\beta_1<\frac{1}{\beta_2}$. The right boundary is the upper-bound on excitation, $\alpha<2\sqrt{\beta_1 \beta_2}$. Simulation results are indicated by colored dots, where a red dot indicates success (contraction) and gray failure. For convenience, only the range $\alpha>1, \beta_1>1$ is shown here.}
\label{figSimulation}
\end{figure}

Do the analytical bounds derived above match the behavior of the system when it is simulated? We simulated a WTA network as described (with 2 excitatory units) and systematically tested different combinations of the parameters $\alpha,\beta_1,\beta_2$. For each simulation we determined whether all units in the network reach steady state with $\dot{x_i}=0$ after a sufficient amount of time. Such networks where classified as stable or unstable, respectively (Fig \ref{figSimulation}). Here, we vary $\alpha$ and $\beta_1$ while keeping $\beta_2=0.25$. While the analytically derived solution is slightly more conservative than necessary it closely matches the results of the simulations (see Fig \ref{figSimulation} and legend for details). The crucial parameter is the excitatory strength relative to the inhibitory strength. This can be seen from the general increase of the permissible value of $\alpha$ as a function of $\beta_1$ (Fig \ref{figSimulation}). Note however that our analytical solution assigns an upper bound to $\beta_1$ as well, which is unnecessary for the numerical simulations. However, strong values of $\beta_1$ lead the system to oscillate and keeping the parameter within the range derived analytically prevents this problem.

\subsection{Stability of single WTA - bump of activity}
\label{sect:alpha2}
The previous analysis considered WTA networks where only $\alpha_1=\alpha>0$ and $\alpha_2=0$ (Fig \ref{figConn}A). In this configuration, the winner of the competition is represented by a single active unit. However, cooperation between neighboring units can also be introduced, by setting $0<\alpha_2<\alpha_1$. The winner is now represented by a more distributed "hill of activity". Our analysis can also be extended to this case. 

For the simplest case of 2 units, this network has the Jacobian
\begin{equation}
\tau \mathbf{J} = \left[ 
\begin{array}{ccc}
l_1 \alpha_1-G & l_1 \alpha_2   & -\ l_1 \beta_1           \\
l_2 \alpha_2   & l_2 \alpha_1-G & -\ l_2\beta_1           \\
l_3 \beta_2    & l_3 \beta_2    & -G                 \\
\end{array} 
\right]
\label{eq:Jalpha2}
\end{equation}
\noindent with $l_{1,2,3}=1$. Using the approach outlined previously, this system is stable if $\mathbf{\Theta}\mathbf{J} \mathbf{\Theta}^{-1}<\mathbf{0}$. After applying the $\mathbf{\Theta}$ coordinate transform, examining the eigenvalues of the Hermitian part of this system reveals that 
\begin{equation}
\frac{1}{2}(-2+\alpha_1+\alpha_2\pm\sqrt{ (\alpha_1+\alpha_2)^2 - 8 \beta_1\beta_2})<0
\label{eq:eigValsAlpha2}
\end{equation}
\noindent is a required condition (plus others, not shown). Comparing this condition to the eigenvalues of the system with $\alpha_2=0$ (see (\ref{eq:eigValsN2})) reveals that $\alpha$ was replaced by $\alpha_1+\alpha_2$ (plus some other minor modifications). This result confirms the intuition that the crucial factor is the total excitatory input $\alpha_1+\alpha_2$ to any one unit. A sufficient condition for this system to be contracting is (compare to Eq \ref{eq:limAlpha})

\begin{equation}
0<\alpha_1+\alpha_2<\sqrt{8\beta_1\beta_2}
\end{equation}
\noindent This condition applies as long as $\alpha_1+\alpha_2<2$ and $\alpha_1-\alpha_2<1$. Together, these conditions similarly permit a fairly wide range of parameters, including $\alpha_1>1$. For example, if $\beta_1=3$, $\beta_2=0.3$ and $\alpha_2=0.5$, $\alpha_1<1.5$.
Note the critical trade-off between the inhibitory gain $\beta_1\beta_2$ and the excitatory gain $\alpha_1+\alpha_2$ that is expressed in this section.

\subsection{Stability of two bidirectionally coupled WTAs}
\label{section:sync}
Next we  consider how two WTAs $x$ and $y$ can be coupled stably (by $\gamma$ connections as shown above). The key idea is first to give sufficient conditions for stable synchronization of the two WTA's. Note that by synchronization we mean here that two variables have the same value (in contrast to other meanings of synchronization i.e. in population coding). This allows the dimensionality of the stability analysis to be reduced. Indeed, synchronization implies that the overall system stability can then be analyzed simply by considering the stability of the individual {\it target} dynamics, i.e., of any one of the subsystems where the external coupling variables have been replaced by the corresponding (endogenous) variables in the subsystem. For instance, in the target dynamics of ${\bf x}$, equation~(\ref{eq:recmapEcoupledT1}) is replaced by
\begin{equation}
\tau \dot{x}_2 + G x_2 = f( I_2 + \alpha x_2 + \gamma x_2 - \beta_1 x_N - T)
\label{eq:targetT1}
\end{equation}
\noindent Next, we shall see that in fact, given the form of coupling we assume, stable synchronization of the subsystems comes ``for free''. That is, it is automatically satisfied as long as sufficient conditions for the stability of the individual target dynamics are satisfied.

Following \cite{PhamSlotine07}, synchronization occurs stably if the following holds:
\begin{equation}
\mathbf{V} \mathbf{J} \mathbf{V}^T < \mathbf{0}
\label{eq:S1}
\end{equation}
\noindent where
\begin{equation}
\mathbf{V} = [ \mathbf{I}_N \ \ -\mathbf{I}_N ]
\label{eq:S2}
\end{equation}
\noindent and $\mathbf{J}$ is the Jacobian of the entire system. 
Here, we define synchrony as equal activity on both maps, i.e. $x_i = y_i$ for all $i$. This condition is embedded in $\mathbf{V}$ as shown. Note that the system need not start out as $x_i = y_i$ to begin with but rather the condition embedded in $\mathbf{V}$ guarantees that the system will converge towards this solution. Other conditions of synchrony (such as only some neurons synchronizing) can similarly be specified by modifying $\mathbf{V}$ accordingly.
$\mathbf{V}$ specifies a metric $\mathcal{M}^{\bot}$  which is orthogonal to the linear subspace $\mathcal{M}$ in which the system synchronizes (i.e. a flow-invariant subspace, see Theorem 3 in \cite{PhamSlotine07}).

The Jacobian $\mathbf{J}$ has dimension $2N$ and is composed of the two sub-Jacobians $\mathbf{J}^1$ and $\mathbf{J}^2$ (as shown in Eq \ref{eq:Jmain}), which describe a single WTA, and of the Jacobians of the couplings.  

\noindent Introducing the coupling term
\begin{equation}
\mathbf{C} = \left[ 
\begin{array}{ccc}
\gamma & 0 & 0 \\
0 & \gamma & 0 \\
0 & 0 & 0 \\
\end{array} 
\right]
\label{eq:C1}
\end{equation}
\noindent results in the Jacobian $\mathbf{J}$ of the full system:
\begin{equation}
\mathbf{J} = \left[ 
\begin{array}{cc}
\mathbf{J}^1 & \mathbf{C}   \\
\mathbf{C}   & \mathbf{J}^2 \\
\end{array} 
\right]
\label{eq:J2}
\end{equation}
\noindent which can be written, using again dummy terms $l_j$, as
\begin{equation}
\tau \mathbf{J} = \left[ 
\begin{array}{cccccc}

l_1 \alpha-G  & 0             & - l_1 \beta_1    & l_1 \gamma   & 0                & 0           \\
0             & l_2 \alpha-G  & - l_2 \beta_1    & 0            & l_2 \gamma       & 0           \\
l_3 \beta_2   & l_3 \beta_2   & -G               & 0            & 0                & 0           \\
l_4 \gamma    & 0             & 0                & l_4 \alpha-G & 0                & - l_4 \beta_1    \\
0             & l_5 \gamma    & 0                & 0            & l_5 \alpha-G     & - l_5 \beta_1    \\
0             & 0             & 0                & l_6 \beta_2  & l_6 \beta_2      & -G          \\
\end{array} 
\right]
\label{eq:J11}
\end{equation}
\noindent The above expression yields:
\begin{equation}
\tau \mathbf{V} \mathbf{J} \mathbf{V}^T = \left[ 
\begin{array}{ccc}
(l_1 + l_4)( \alpha - \gamma)- 2G            &            0                                  & -\beta_1 (l_1+l_4)  \\ 
                        0                    &  (l_2+l_5) (\alpha - \gamma) - 2G               & -\beta_1( l_2+l_5)   \\
 \beta_2 (l_3+l_6)                           &            \beta_2 (l_3+l_6)                  &  -2G                  \\
\end{array} 
\right]
\label{eq:VJV}
\end{equation}
\noindent Note that $\alpha>0, \beta_1>0, \beta_2>0$. 

Consider now the Jacobian of e.g. subsystem-1 once synchronized, i.e.,
with the coupling terms from subsystem-2 variables replaced by the same
terms using subsystem-1 variables (this is what we called earlier the target
subsystem-1). Given equation (\ref{eq:Jmain}) and (\ref{eq:targetT1}), this Jacobian can be
written
\begin{equation}
\tau \mathbf{J}^1_{sync} = \left[ 
\begin{array}{ccc}
l_1 (\alpha+ \gamma)-G & 0        & -\ l_1 \beta_1           \\
0        & l_2 (\alpha+ \gamma)-G & -\ l_2\beta_1           \\
l_3 \beta_2  & l_3 \beta_2  & -G                 \\
\end{array} 
\right]
\label{eq:J1_sync}
\end{equation}
\noindent Comparing (\ref{eq:VJV}) and (\ref{eq:J1_sync}), we see that sufficient
conditions for $\mathbf{J}^1_{sync}$ (and similarly
$\mathbf{J}^2_{sync}$) to be negative definite automatically imply
that $\mathbf{V} \mathbf{J} \mathbf{V}^T$ is negative
definite. Indeed, since $\gamma \ge 0$,
\begin{equation}
\forall \ l_j\ ,\ \mathbf{J}^1_{sync} < \ {\bf 0}\ \ \ \ \ =>\ \ \ \ \ \forall \ l_j\ ,\ \mathbf{V} \mathbf{J} \mathbf{V}^T < \ {\bf 0}
\label{eq:JsyncVJV}
\end{equation}
\noindent In other words, the basic requirement that the individual
target dynamics are stable (as shown in the previous section) automatically implies stability of the
synchronization mechanism itself.

Note the opposite signs of $\gamma$ in Eqs (\ref{eq:J1_sync}) and
(\ref{eq:VJV}). Intuitively, these express a key trade-off. Indeed,
the stronger $\gamma$ is, the easier and stronger the synchrony of the
memory state (Eq (\ref{eq:VJV})). However, a stronger connection also
makes the system less stable. This is expressed by the positive
$\gamma$ in Eq (\ref{eq:J1_sync}), which imposes stricter constraints on
the permissible values of the other weights for $\mathbf{J}^1_{sync}$
to remain negative definite.

Synchronization of the two maps in this way allows reduction of the two coupled systems to a single virtual system with the additional parameter $\gamma$ for the coupling strength (Eq \ref{eq:VJV},\ref{eq:J1_sync}). Stability of this hybrid system guarantees stability of the synchronization mechanism itself (Eq \ref{eq:JsyncVJV}). The upper-bounds for $\gamma$ are thus (based on Eq \ref{eq:limAlpha})

\begin{equation}
\gamma<2 \sqrt{\beta_1 \beta_2} - \alpha
\label{eq:keyResGamma}
\end{equation}
\noindent As long as this condition is met, the dynamics of each map are contracting and their synchronization is stable. The lower-bound on $\gamma$ is determined by the minimal activity necessary to begin "charging" the second map (which gets no external input in our configuration). The minimal activity that a unit on the second map gets as input from the first map needs to be larger than its activation threshold $T$, i.e. $x_i \gamma > T$ where $x_i$ is the steady-state amplitude during the application of input (which is a function of the gain $g=\frac{1}{1+\beta_1\beta_2-\alpha}$). Thus, 

\begin{equation}
\frac{T}{g I_{max}(t)} < \gamma 
\end{equation}

\subsection{Stability of unidirectionally coupled WTAs}
Next, we extend our analysis to networks consisting of 3 WTAs  $\mathbf{x}$, $\mathbf{y}$ and $\mathbf{z}$ of the kind shown in Fig \ref{figConn}D and described in section \ref{sec:undirect}. WTAs $\mathbf{x}$ and $\mathbf{y}$ are bidirectionally coupled to express the current state and are equivalent to the network considered in the previous sections. A further WTA $\mathbf{z}$ is added that contains units $z_i$, referred to as transition neurons (TNs). In this example, there are two TNs $z_1$ and $z_2$ (Fig \ref{figConn}D). Activation of the first ($z_1$) leads the network to transition from state $x_1$ to $x_2$, if the network is currently in $x_1$. Activation of the second ($z_2$) leaves the network in state $x_2$, if the network is currently in this state. If it is not so, then no activity is triggered. The TN $z_1$ is an example of a transition from one state to another. TN $z_2$ is an example of a transition that starts and ends in the same state (a loop). This loop is intentionally introduced here, because it poses a limit to stability. TNs receive and project input with weight $\phi>0$.

The Jacobian of the full system consists of $3N$ variables:
\begin{equation}
\tau \mathbf{J} = \left[ 
\begin{array}{ccc}
\mathbf{J}^1 & \mathbf{C}   & \mathbf{P}^2 \\
\mathbf{C}   & \mathbf{J}^2 & \mathbf{0}   \\
\mathbf{0}   & \mathbf{P}^1 & \mathbf{J}^3 \\
\end{array} 
\right]
\label{eq:J_TN}
\end{equation}
\noindent Since there are two memory states, 
\begin{equation}
\mathbf{C} = \left[ 
\begin{array}{ccc}
\gamma & 0      & 0 \\
0      & \gamma & 0 \\
0      & 0      & 0 \\
\end{array} 
\right]
\label{eq:C1_TN}
\end{equation}
\noindent $\mathbf{P}^1$ describes the input and $\mathbf{P}^2$ the output of the TNs. Here,
\begin{equation}
\mathbf{P}^1 = \left[ 
\begin{array}{ccc}
\phi   & 0    & 0 \\
0      & \phi & 0 \\
0      & 0    & 0 \\
\end{array} 
\right]
\label{eq:P1}
\end{equation}
\begin{equation}
\mathbf{P}^2 = \left[ 
\begin{array}{ccc}
0      & 0    & 0 \\
\phi   & \phi & 0 \\
0      & 0    & 0 \\
\end{array} 
\right]
\label{eq:P2}
\end{equation}

\subsubsection{Case 1: loop}
For purposes of worst-case analysis, assume that the TN $z_2$ (which receives and projects to state 2)
is permanently active. This is achieved by setting $T_{TN}=0$. In this case, we require that the network remains synchronized in state $z_2$. 

The state is stable with $z_2$ activated if the synchrony between $\mathbf{x}$ and $\mathbf{y}$ is not disrupted. This is the case if
\begin{equation}
\tau \mathbf{V} \mathbf{J} \mathbf{V}^T = 
\left[ 
\begin{array}{ccc}
(l_1+l_4)(\alpha-\gamma) -2 &     0    & -\beta_1(l_1+l_4) \\
0                  & (l_2+l_5) (\alpha  - \gamma)-2 & -\beta_1(l_2+l_5) \\
\beta_2 (l_3+l_6)  & \beta_2 (l_3+l_6)  & -2 \\
\end{array} 
\right]
\label{eq:VJV_TN}
\end{equation}
\noindent with $ \mathbf{V} = \left[\mathbf{I}_N \ \ -\mathbf{I}_N \ \ \ \mathbf{0}_N \right]$
and $\mathbf{J}$ the Jacobian of the entire system (9 variables).

Note the similarity to equation~(\ref{eq:VJV}). None of the non-linearity terms of the 3rd WTA $l_7,l_8,l_9$, nor $\phi$, appear in this equation. Thus, the synchrony of the states is not influenced by the presence of a consistently active loop TN. The presence of $\phi$ does thus not influence the synchrony between $x$ and $y$ which represent the state. However, this combined system also needs to be contracting for this system to be stable (i.e. reach steady state). Thus, we next derive the limits on $\phi$ for this to be the case.

Using the insight gained in section \ref{section:sync}, we replace the $y_i$ terms by $x_i$ terms for purposes of stability analysis. Note that the principle of showing synchronization first introduces a hierarchy (or series) of dynamic systems, so that the overall result converges if each step (sync and simplified system) does, with convergence rate the slowest of the two. In our case the synchronization step is always the fastest, so the overall convergence rate is that of the reduced system.

Next, we analyze the stability of the reduced system (consisting of $\mathbf{x}$ and $\mathbf{z}$). Here, only $z_2$ (loop TN) is
used, $z_1$ is not connected. The corresponding Jacobian is:
\begin{equation}
\mathbf{J^{TN}} = \left[ 
\begin{array}{cccccc}
l_1 (\alpha+\gamma)-1  & 0                      & - l_1 \beta_1    & 0            & 0                & 0                \\
0             & l_2 (\alpha+\gamma)-1  & - l_2 \beta_1    & 0            & l_2 \phi       & 0                 \\
l_3 \beta_2   & l_3 \beta_2            & -1               & 0            & 0                & 0                 \\
0             & 0                      & 0                & l_7 \alpha-1 & 0                & - l_7 \beta_1     \\
0             & l_8 \phi               & 0                & 0            & l_8 \alpha-1     & - l_8 \beta_1     \\
0             & 0                      & 0                & l_9 \beta_2  & l_9 \beta_2      & -1                \\
\end{array} 
\right]
\label{eq:JredTN}
\end{equation}
\noindent Having $\mathbf{J}^{TN}$ be negative definite in the metric $\mathbf{\Theta}$,
\begin{equation}
\forall \ l_j\ ,\ \mathbf{\Theta} \mathbf{J}^{TN} \mathbf{\Theta}^{-1}< \ {\bf 0}\ 
\label{eq:JredTN_stab}
\end{equation}
guarantees that the coupled system is stable. Following  \cite{WangSlo} and \cite{Slotine03} (section 3.4), if the uncoupled systems are stable with contraction rates $\lambda_x$ and $\lambda_z$, then the coupled system is stable if
\begin{equation}
\phi^2 \ < \ \lambda_x \lambda_z
\label{eq:phi2}
\end{equation}
and its contraction rate is
\begin{equation}
\lambda_{x,z} = \frac{(\lambda_x + \lambda_z)}{2} - \sqrt{ (\frac{\lambda_x - \lambda_z}{2})^2 + \phi^2}
\end{equation}
Note that $\ \lambda_{x,z} > 0$ is equivalent to condition (\ref{eq:phi2}). One then has $0 \ < \ \lambda_{x,z} \ \le \ \lambda_x \ \le \ \lambda_z$. Note that if the connection weights are not symmetric, $\phi$ in the expressions above can be replaced by $\phi=\frac{\phi_1+\phi_2}{2}$.

The contraction rate for a single WTA is equal to the absolute value of the largest eigenvalue of $\mathbf{F}_s$ (its real part) \cite{WangSlo}. Following (\ref{eq:TW2}), the contraction rate for a WTA (such as $\mathbf{z}$) is 
$\lambda_{z} = | Re( \frac{1}{2}(-2+\alpha+\sqrt{\alpha^2-4\beta_1\beta_2}) )|$.
Similarly, for a symmetrically coupled system with coupling weight $\gamma$, the contraction rate is
$\lambda_{x} = | Re( \frac{1}{2}(-2+\alpha+\gamma+\sqrt{(\alpha+\gamma)^2-4\beta_1\beta_2}) )|$.
These two conditions thus establish the upper-bound on the permissible weight of $\phi<\sqrt{\lambda_x\lambda_z}$. Since $\lambda_x<\lambda_z$, a good approximation is

\begin{equation}
\phi<\lambda_x
\label{eq:keyResPhi}
\end{equation}

\subsubsection{Case 2: transition}
Here, the transition from one pattern of synchrony to an other (two states) is investigated. For this purpose, both states $x_1$ and $x_2$ exist. Also, the TN $z_1$ is connected. Since activating $z_1$ leads from a transition from state 1 to 2 (represented by $x_1$ and $x_2$). In the following, we assume that the network is in $x_1$ when initialized and $z_1$ is active (i.e. $T_{TN}=0$). We then show that the network will end up in the second synchrony pattern, representing $x_2$. 

Defining ${\bf V}$ as above and $\mathbf{J}$ the appropriate Jacobian of the full system yields
\begin{equation}
\tau \mathbf{V} \mathbf{J} \mathbf{V}^T =
\left[ 
\begin{array}{ccc}
(l_1+l_4) (\alpha - \gamma) -2 &     0    & -\beta_1(l_1+l_4) \\
0                  & (l_2+l_5) (\alpha  - \gamma) -2 & -\beta_1(l_2+l_5) \\
\beta_2 (l_3+l_6)  & \beta_2 (l_3+l_6)  & -2 \\
\end{array} 
\right]
\label{eq:VJV_TN2}
\end{equation}
\noindent Similarly to equation~(\ref{eq:VJV_TN}), the terms of the 3rd WTA do not appear. Thus, activation of the TN does not disturb the synchrony between $\mathbf{x}$ and $\mathbf{y}$ as such but only which particular units synchronize (this is not visible in above equation).

Whether or not the system transitions from one pattern of synchrony to another is determined by the threshold behavior of the activation function. As long as the input to $x_2$, $\phi z_1>0$ for sufficient amount of time, the network will switch its pattern of synchrony. If, on the other hand, $z_1$ is not active for a sufficiently long time (relative to the contraction rate $\lambda_x$), the system will return to the previous pattern of synchrony. Also note that $z_1$ switches off automatically as soon as the transition occurs (since then $y_1$ ceases to be active). Thus, the timing of the external input does not have to be tightly connected with the external dynamics, or can even be permanently present.

\begin{figure}
\centering
\includegraphics[angle=0,width=11cm]{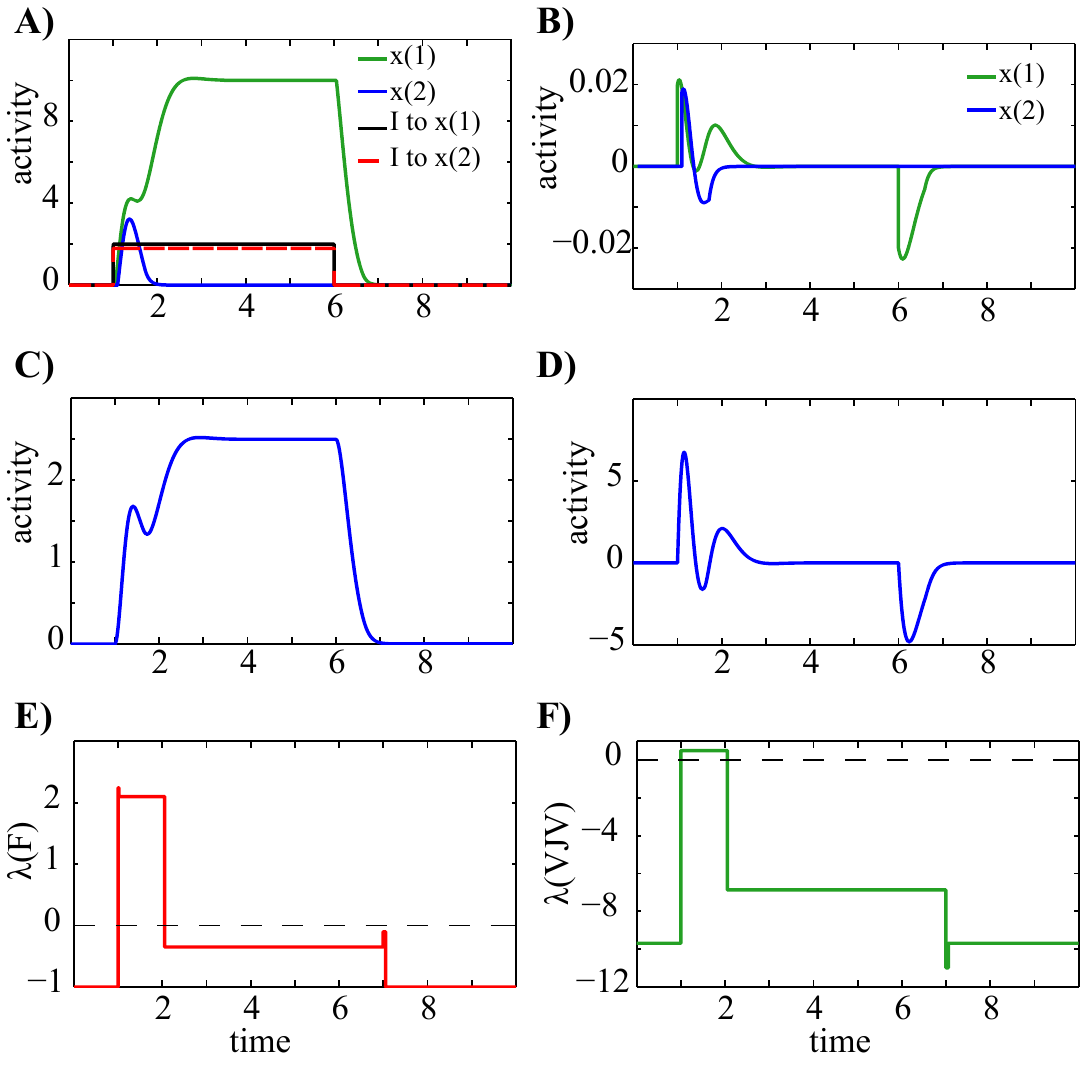}
\caption{Simulation of a single WTA network (Fig \ref{figConn}B). 
(A) Excitatory units and external input. 
(B) First derivative with respect to time for the excitatory units. The activity of $x_2$ is offset by 100 timesteps relative to $x_1$ for plotting purposes only.
(C) Activity of the inhibitory unit. 
(D) Derivative of the inhibitory unit. 
(E) The maximal eigenvalue $Re(\lambda_{max}(F_s))$ of the generalized Jacobian. 
(F) The minimal eigenvalue of $\lambda_{min}(VJV_{s})$. 
Activity is plotted in arbitrary units, x-axis is in units of integration timesteps in units of 1000 (Euler integration with $\Delta=0.01$).  See the text for details.
}
\label{figSim1}
\end{figure}

\begin{figure}
\centering
\includegraphics[angle=0,width=11cm]{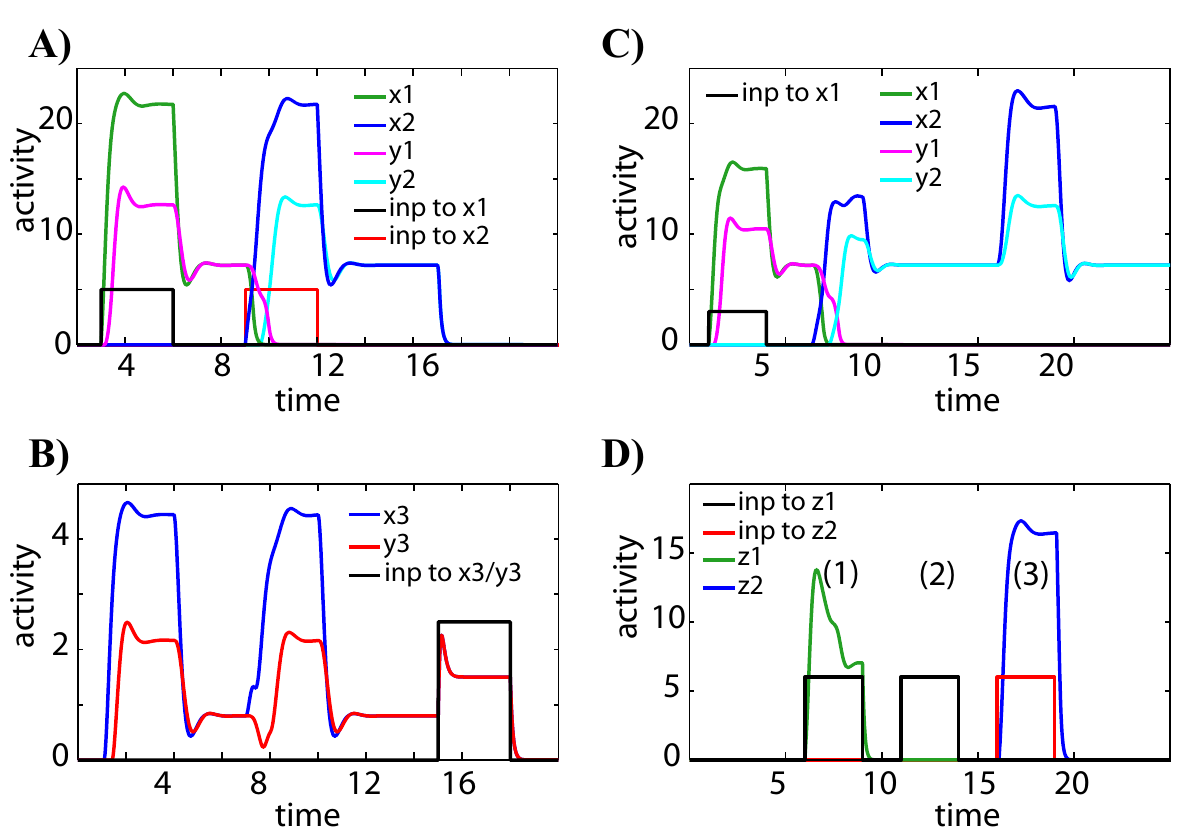}
\caption{Simulation of networks consisting of two (A-B) and three (C-D) recursively coupled WTA networks.
(A-B) Simulation of two symmetrically coupled WTAs as shown in Fig \ref{figConn}D (with $\phi=0$). In this network, there are two $\gamma$ connections between the maps (between $x_1,y_1$ and $x_2,y_2$). (A) shows the excitatory units, (B) the inhibitory units. 
(C-D) Simulation of three coupled WTAs, as illustrated in Fig \ref{figConn}D. Maps $x$ and $y$ are coupled symmetrically as in (A-B), whereas map $z$ creates a unidirectional feedback loop between $y$ and $x$ (see Fig \ref{figConn}D for details). Shown is the activity of the excitatory units on $x$ and $y$ (C) as well as on the third map ($z$, panel D). Units of time are integration timesteps in units of 1000. See text for details.
}
\label{figSim2}
\end{figure}

\begin{figure}
\centering
\includegraphics[angle=0,width=11cm]{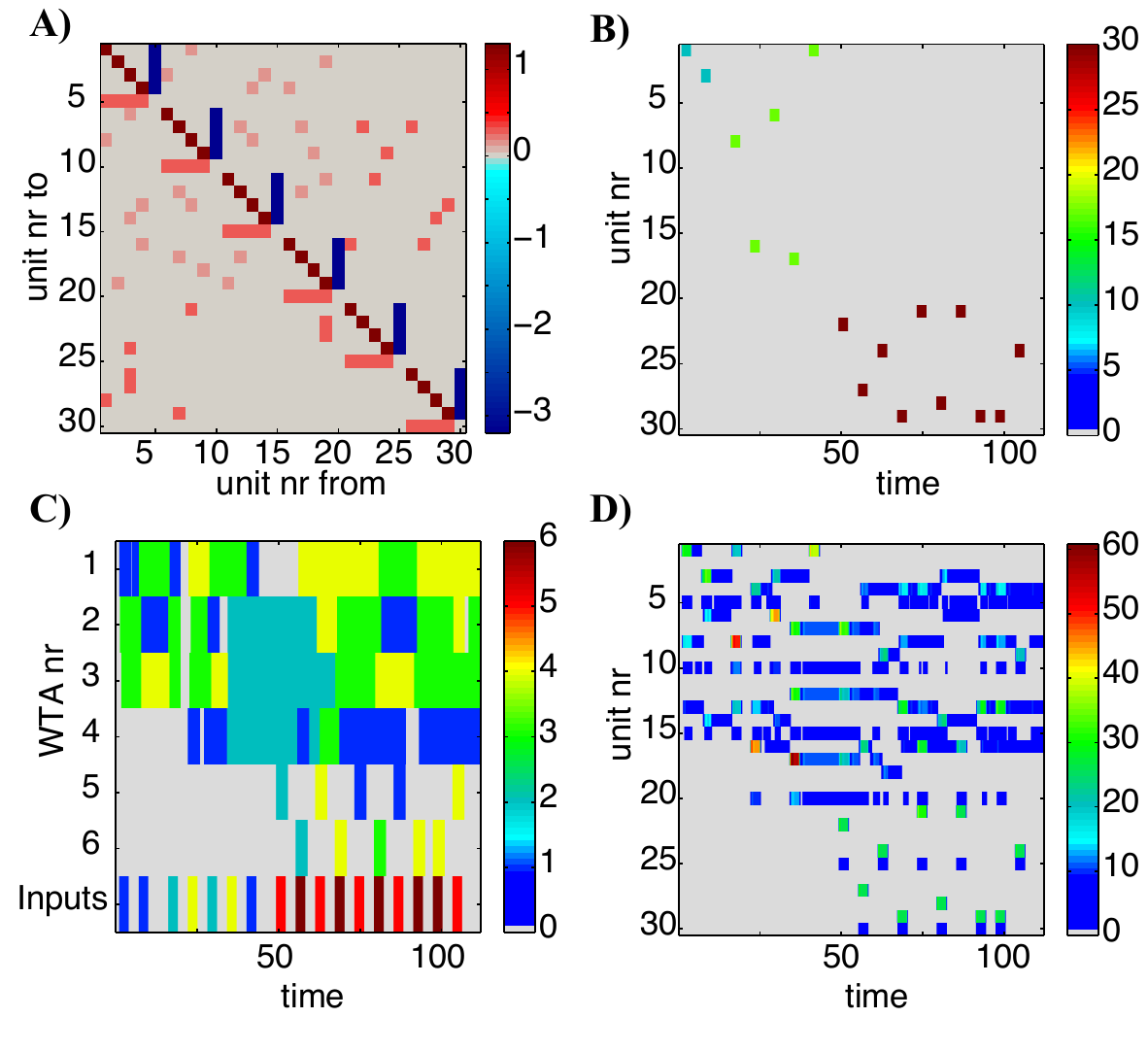}
\caption{Simulation of a randomly connected network consisting of 6 WTA modules. 
(A) Weight matrix of the entire network. (B) External inputs used for the simulation shown in C-D. (C) Color-coded diagram of the winner on each WTA. Winning units are either none (0) or 1-4 (indicated by the color, representing 0-4). The bottom row shows to which WTA external input is currently provided (1-6). (D) Plot of the activity of every unit in the network during application of the input shown in B.  Note how on every WTA, only one unit can be active (after convergence). The color-code represents the activity of each unit, in arbitrary units. Units of time are integration timesteps in units of 1000.
}
\label{figSim3}
\end{figure}

\subsection{Verification by simulation}

\subsubsection{Single WTA}
The properties of contraction during the application of external input can be demonstrated by numerical simulation of a single WTA network consisting of 2 excitatory and 1 inhibitory units (see Fig \ref{figConn}B for wiring). Figure \ref{figSim1} shows the dynamics of the 3 units in the network while external input is applied to the two excitatory units $x_1$ and $x_2$. The input $I(t)$ is a binary pulse of amplitude 2.0 to $x_1$ and 1.8 to $x_2$ (difference 10\%). Note how the network amplifies the small difference ($x_1$ is the winner). Parameters were: $\alpha=1.3, \beta_1=2, \beta_2=0.25, T=0, G=1$, i.e. the gain was $g=5$ (steady-state amplitude $g*I$). These parameters satisfy all conditions for contraction. The properties of contraction are evaluated at every point of time by evaluating the effective Jacobian at this point of time. The maximal eigenvalue $Re(\lambda_{max}(F_s))$ of the generalized Jacobian indicates whether the network is currently contracting or not. Whenever the values is below zero (dashed line), $F_s$ is negative definite.
 
The minimal eigenvalue of $\lambda_{min}(VJV_{s})$ (Fig \ref{figSim1}E) indicates points of time when the network is not contracting. Whenever the value is above the dashed line, $VJV^T$ is positive definite. Note the interplay of the dynamics of the maximal eigenvalue of the generalized Jacobian and the minimal eigenvalue of $VJV^T$, which together reflect the dynamic state of the system. Whenever the system is contracting, $\lambda_{max}(F_s)<0$. Shortly after the onset of the external input (t=1000), no winner has been selected and the system is not contracting. Instead it is diverging exponentially towards a contracting region of the state space. Note also the other important transition of the system after the offset of the input (t=6000). After a while only the inhibitory neuron is active, which explains the change around t=7000.

\subsubsection{Two-and three WTA networks}
Next we simulated two networks: one consisting of two (Fig \ref{figSim2}A-B) and one consisting of 3 (Fig \ref{figSim2}C-D) recursively coupled WTAs, connected by either bidirectional or directional connections. Parameters were: $\alpha=1.3, \beta_1=2.8,\beta_2=0.25, \gamma=0.15, T=1$. For the second simulation, in addition $T_{TN}=5, \phi=0.3$. These simulations (see the legend of Fig \ref{figSim2} for details) illustrate the dynamics of the network during persistent state dependent activity that exists due to the $\gamma$ connections. Also, it shows how these states can be utilized to implement state dependent transitions (see Fig \ref{figConn}D for connectivity). 

Notice how, after application of external input to $x_1$, activity persists (Fig \ref{figSim2}A). After applying input to $x_2$, the winner switches. Increasing inhibition (onset at t=15000, panel B) globally resets both maps. 
Note also how application of external input to $z_1$ leads to a transition from the first to the second state only if the network is in state $x_1$ (case (1) vs (2) as indicated in Fig \ref{figSim2}D). This illustrates a state dependent reaction of the network to external input. The third input application (case (3) in Fig \ref{figSim2}D) illustrates the stability of the network, even if the transition is onto itself. Note how the network reaches a stable level of activity before the external input to $z_2$ is removed, indicating balanced inhibition/excitation despite multiple feedback loops both within and between maps. 

\subsubsection{Large networks}
Contraction properties are particularly useful because they are preserved if several contracting systems are combined to form a larger system \cite{Slotine03,LohSlo98}. Such systems can be combined in several ways, including parallel combinations, hierarchies, and certain types of feedback to form a new system. The resulting composite system is guaranteed to be contracting as well if the underlying sub-systems are contracting \cite{Slotine01,Slotine03}. Note that, in themselves, combinations of stable modules have no reason to be stable \cite{Slotine01}. This is only guaranteed if the constituting modules are contracting. To illustrate that contracting WTA networks can be used to construct larger networks, we next simulated a network consisting of 6 identical WTAs (Fig \ref{figSim3}). The purpose of this simulation is to demonstrate that satisfying contraction-properties at the level of the constituting modules (one WTA) is sufficient to guarantee stability of a large connected network with many (potentially unknown) feedback loops. 

Each WTA consists of 4 excitatory and 1 inhibitory unit. The first 4 WTAs represent states and remaining two state-dependent transitions. Bidirectional $\gamma$ connections were generated randomly between the excitatory units of these 4 WTAs. Unidirectional $\phi$ connections where placed randomly between units on the last 2 WTAs and units on the first 4 WTAs. Parameters were $\alpha=1.3$, $\beta_1=3.2$, $\beta_2=0.25$, $\gamma=0.15$, $T=1$, and $\phi=0.3$. Inputs were generated randomly, first restricted to only the state-carrying WTAs and later only to the transition-inducing units. One instance of this simulation is shown in Fig \ref{figSim3}. Note the following features of the result: i) the network is in a non-linear hard-WTA configuration: except for transients, only one excitatory unit in each WTA is active (Fig \ref{figSim3}D). ii) The reaction to the same external input depends on the state of the network. For example, consider the inputs to WTA 5 in Fig \ref{figSim3}C. Input to unit 1 (blue) is provided twice, but only once does this induce a state switch (visible as change of winners on the first WTAs). iii) levels of activity are strictly bounded, despite external input and high gain of the network (Fig \ref{figSim3}D). Similar simulations with other weights (even randomized) are also stable, as long as the parameters are within the ranges permitted.

Simulation of a networks consisting of 100 and 1000 WTAs, probabilistically connected as described above, were found to be stable as well. We assume the parameters of the WTA itself are static. The critical concerns are the connections between the WTAs ($\gamma$ and $\phi$). As long the pairwise connection probability between a given WTA and all other WTAs is sufficiently low, stability will be guaranteed. This is because the sum of all $\gamma_j$ which connect to a particular unit needs to observe the constraint shown in Eq \ref{eq:keyResGamma}: $\sum_{j} \gamma_j \leq 2\sqrt{\beta_1\beta_2}-\alpha$.

\section{Discussion}
Biology is replete with examples in which relatively independent sub-systems are coupled together by various degrees of time-varying positive and negative feedback, and nevertheless the entire system is functionally stable. The neocortex, with its billion neurons parcellated into millions of interconnected local circuits is one striking example of such a system. We have chosen to study this cortical case, because the functional stability of the vastly interconnected cortical system is intimately associated with the expression of stable intelligent behavior. 

Our contribution toward understanding this intriguing phenomenon of collective stability, is that we have identified and properly analyzed
a small functional module (here a WTA circuit) and showed that this knowledge alone allows us to guarantee stability of the larger
system. By instability we here mean anything that leads to run-away behavior, which in
biological neurons means the neurons will latch into saturation and, if active long enough, die \cite{Syntichaki03}.
We do not consider this seizure-like pathological state to offer 'boundedness' in any useful sense. Instead, it is the case that neurons usually operate at a small fraction of their maximum firing rate, and their networks are only stable because of tight inhibitory-excitatory balance. The brain pays considerable attention to maintaining this balance. For these reasons the stability of neuronal networks is better cast as stability in a network of (effectively) positively unbounded linear threshold neurons, than for example sigmoidal neurons. The simplified model neurons we used make the problem tractable, while preserving the features of real neurons that we believe are crucial to understand neuronal circuits and their stability. We thus expect our observations to be directly applicable to spiking neurons. 

The strength of these results is that they cast light on the problem of how global stability can be achieved in the face
of the locally strong feedbacks required for active, non-linear signal processing. We have examined this problem in networks of WTA
circuits, because the WTA is a rich computational primitive, and because neuroanatomical connectivity data suggest that WTA-like
circuits are a strong feature of the networks formed by neurons in the superficial layers of cortex.

In essence, we have employed Contraction Analysis to demonstrate that a WTA network can at the same time be contracting and strongly amplifying. Our key results are the bounds documented in Eqs \ref{eq:keyResultBegin}-\ref{eq:keyResultEnd}, \ref{eq:keyResGamma}, \ref{eq:keyResPhi}  and illustrated in Fig \ref{figGraphicalIneq}. It is important to note that this analysis could not have been performed with standard linear tools (such as eigenvalues of Jacobians at fixed points), which rely on linearizations and do not provide global stability conditions. 
While in principle the asymptotic convergence may have been demonstrated using a Lyapunov function \cite{SlotineLi91}, actually no such function is known for these kinds of networks. 
In contrast, using contraction analysis we could demonstrate exponential (as opposed to asymptotic) convergence for arbitrary initial conditions.
In addition to systematic analysis, we have also confirmed our results with simulation of random systems composed of WTAs. The WTA network thus constitutes a strong candidate for a canonical circuit which can serve as a basis for the bottom-up construction of large biological and artificial computational systems. Our approach is similarly applicable to functional modules other than WTAs, such as liquid state machines with feedback \cite{Maass07}.

To systematically analyze subsets of active neurons which should either be contracting or not (i.e. permitted winners or not) we utilized what we called $l_i$ terms. The parcellation of the active subsets of the network using such terms can be regarded as a generalization of the approach of permitted and forbidden sets ("switching matrix") \cite{Hahnloser03}. However, that previous approach is suitable only for fully symmetric networks, whereas the proper operation of a WTA network requires asymmetry of the inhibitory connections. Our approach is to exhaustively show for any possible subset of the $l_i$ terms that it is either exponentially diverging or contracting. This way, the network as a whole is guaranteed to exponentially converge to one of the fixed points. The concept we developed can be applied at any point of time during the operation of the network. In particular it can be applied before the winner is known. Our reasoning indeed guarantees that a winner will be selected exponentially fast.

Winner-take-all networks are representatives of a broad class of networks, where a number of excitatory units share a common inhibitory signal that serves to enforce competition \cite{AbbottDayan01,Amari77b,YuilleGeiger03,Hertz91,TankHopfield86,Rabinovich00,Ermentrout92,Schmidhuber89}. There are many instances of networks that share this property, including various neural networks but also gene regulatory networks, in-vitro DNA circuits \cite{KimHopfieldWinfree04} and development. Competition enforced by shared inhibition among excitatory units is a principal feature of brain organization \cite{Kurt08,Baca08,Tomioka05,Pouille2009,Buzsaki84,Mittmann2005,Gruber09,Sasaki07,Papadopoulou10} and our findings are thus directly applicable for reasoning about such circuits. WTA-type behavior has been experimentally demonstrated in a variety of brain structures and species including the mammalian hippocampus and cortex \cite{Kurt08,Baca08,Mittmann2005,Gruber09,Sasaki07,Busse09}. Also, the existence of functional WTA circuits has been suggested based on strong anatomical evidence in others, in particular 6-layer cortex \cite{Binzegger04,Tomioka05}. 

Combining several WTA networks permits the implementation of computational operations which can either not be performed by a single WTA or which would require an unrealistically large WTA. In this paper, we show the constraints that need to be imposed on combinations of such WTAs such that the new combined system is guaranteed to be stable.  As an illustration of the computational ability of a network consisting of several WTAs, we simulated a network of 3 WTAs coupled both using symmetric and asymmetric connections (Figure \ref{figSim2}). This network demonstrates two crucial properties that combinations of WTAs permit: persistent activity in the absence of input and state-dependent reaction to external input. While originally designed to demonstrate the stability property, this is also a novel generalization of our previous state-dependent processing network \cite{RutishauserDouglas2009} as well as the pointer-map approach \cite{Hahnloser99}. Note that in contrast to the previous work, here all 3 WTAs are fully homogeneous (identical). The only modifications needed are to establish appropriate connections between the WTAs. In contrast, our previous networks consisted of one or several WTAs plus specialized units. This new and more generic version makes the networks more stable, as excitatory units can only exist as part of a WTA and thus always receive balanced inhibitory input. As in the original \cite{RutishauserDouglas2009}, this enhanced three WTA network is also capable of implementing any regular language (a state automaton).

While we demonstrated our approach for WTAs, our approach is sufficiently general to reason systematically about the stability of any network, biological or technological, composed of networks of small modules that express competition through shared inhibition. For example, synthetic DNA circuits can perform computations, self-assemble and provide a natural way to enforce competition through shared inhibition \cite{KimHopfieldWinfree04,Rothemund04,Adleman94}. Both natural and synthetic gene regulatory networks depend on networks of stochastic chemical reactions, resulting in a system of many nested feedback loops. Robustness is thus a crucial issue \cite{Soloveichik09} and the notion of contracting modules to describe complex aggregates might provide crucial insights into such systems. These networks can also implement WTA-like computations \cite{KimHopfieldWinfree04}. Self-assembly of synthetic DNA circuits and biological tissue in general relies on de-novo bottom-up construction, analogous to our notion of first building small contracting modules and then composing a system consisting of such modules. 

We expect that our results will have immediate application in interpreting the behavior of biological neural networks. For example, most synapses in the nervous system are both unreliable and plastic and so the postsynaptic effect elicited by given action potential is uncertain and time varying. Contracting systems and their aggregations remain stable under these constraints provided the parameters remain within certain well-defined and rather broad ranges. Time-varying changes both in the input as well as the structure of the network is thus permitted within a broad range without endangering the stability of the system \cite{LohSlo98}. This is particularly important if the system is modifying itself through processes such as developmental growth, synaptic plasticity or adult neurogenesis. 

A key question in neuroscience is how a large system such as the brain can rapidly modify itself to adapt to new conditions. A possible solution that our results suggest is that some parts of the network (the modules, here WTAs) could be pre-specified whereas the connections between the modules are learned or modified. This would greatly reduce the required amount of learning or developmental growth processes. A key question is how rules of plasticity can be utilized to enable such learning. It is also an open question whether and how a given WTA can be systematically decomposed into a combination of smaller WTAs that still perform the same function. This question is crucial because the realistic size of a given WTA is restricted by the size of the projection field of recurrent inhibition. Our results provide a framework for investigation of these important questions.

\section{Appendix}
\subsection{Appendix A: Constraints for persistent activity}
The memory state requires $\dot{x}_i = 0$ while $x_i>0$ for the unit $i$ that represents the last winner. At this point, some necessary constraints on the weights can be derived. Since $x_i>0$, unit $i$ is fully linear. The activity of this unit is described by

\begin{equation}
\tau \dot{x}_i = x_i (\alpha-G) - \beta_1 x_N - T_i = 0
\end{equation}

Thus, $x_i (\alpha-G) = \beta_1 x_N + T_i$. It follows that $\alpha>G$ for $x_i$ to be positive. Intuitively, $\alpha-G$ represents the
effective recurrent input the unit receives after accounting for the load (which causes it to decay exponentially to zero
in the absence of input). This effective recurrent input needs to be strong enough to account for the negative inhibitory
input as well as the constant current subtracted by the threshold.  Also, in the two-map case, $\gamma$ can be incorporated in this argument as $\alpha+\gamma>G$ (this follows from the synchrony, see below).  Experimental measurements in neocortex indicate that $\alpha>G$ \cite{Douglas95}. The open-loop gain of such systems is $>1$ (sometimes substantially so). Thus, these networks are only stable because of balanced inhibition. The requirement to operate in the $\alpha>G$ posses strict requirements for stability.

\subsection{Appendix B: Notes on approximation of the activation function}
To motivate the $l_j$ terms, consider the smooth non-linearity $f(x)=log(a+exp(b (x+c))$ with $a=1$, $b=1$ and $c=0$.  Then, $\dot{x}=\frac{exp(x)}{exp(x)+1}$. For large $x$, this is approximately equal to $1$ and for negative $x<0$ approximately equal to $0$. An approximate solution to the derivative is thus either $0$ or $1$. Indeed, the sufficient stability conditions provided by the following analysis will be unchanged if the nonlinearity is replaced by a general sigmoid of slope within the interval $[0, 1]$, corresponding to the dummy terms taking {\it any} time-varying value between $0$ and $1$. In optimization, a frequently used approximation is $p(x,a)=x + \frac{1}{\alpha} \log(1+\exp( -a x))$ (the integral of the sigmoid function) \cite{ChenMangasarian95}. $a>0$ is a constant. Its first derivative is $\frac{1}{1+exp(-a x)}$ and its second derivative is $\frac{a \exp(-a x)}{(1+\exp(-a x))^2}$. Its derivatives are bounded between 0 and 1 and this function can thus be used as a approximation. The approach that we use in this paper is thus similarly valid for this and other smooth approximations (as long as the derivative is bounded $<=1$).

\subsection{Appendix C: Time-constants}
All parameters given in the paper assume that the time-constant $\tau$ is the same for inhibitory and excitatory units. Similar conditions can be derived if this is not the case. In particular, it is of biological interest to consider inhibitory time-constants $\tau_{I}$ which are larger than the excitatory time-constants $\tau_{E}$. Taking possibly different time-constants into account, the bounds in Eqs \ref{eq:keyResultBegin}-\ref{eq:keyResultEnd} become

\begin{equation}
1<\alpha<2 \sqrt{\beta_1 \beta_2 \frac{\tau_E}{\tau_I}}+1-\frac{\tau_E}{\tau_I}
\label{eq:keyResultBeginWithTau}
\end{equation}
\begin{equation}
\frac{1}{4} \frac{\tau_E}{\tau_I} < \beta_1\beta_2 < \frac{\tau_E}{\tau_I}
\end{equation}
\noindent Note that the key variable is the ratio $\frac{\tau_E}{\tau_I}$. If $\tau_E=\tau_I$, the bounds reduce to Eqs \ref{eq:keyResultBegin}-\ref{eq:keyResultEnd}. 

\section{Acknowledgements} 
We acknowledge funding by the California Institute of Technology, the Massachussets Institute of Technology and the EU DAISY and SECO grants. We thank Emre Neftci for discussion and anonymous reviewers for useful suggestions.

\end{document}